\documentclass[twocolumn,notitlepage,prb,superscriptaddress,showpacs,floatfix]{revtex4-1}
\usepackage{hyperref}
\usepackage[usenames,dvipsnames]{color}
\usepackage{amsmath}
\usepackage[english]{babel}
\usepackage{amssymb}
\usepackage{graphicx}
\usepackage{epstopdf}
\usepackage{ulem}
\usepackage{lineno}
\usepackage{tabularx}
\usepackage{float}

\begin{document}
\title{Diagrammatic Monte Carlo for Dual Fermions}

\author{Sergei Iskakov}
\email{iskakoff@q-solvers.ru}
\affiliation{Department of Physics, University of Michigan, Ann Arbor, Michigan 48109, USA} 
\author{Andrey E. Antipov}
\affiliation{Department of Physics, University of Michigan, Ann Arbor, Michigan 48109, USA} 
\author{Emanuel Gull}
\affiliation{Department of Physics, University of Michigan, Ann Arbor, Michigan 48109, USA}

\date{\today}

\begin{abstract}
We introduce a numerical algorithm to stochastically sample the dual fermion perturbation series around the dynamical mean field theory, generating all
topologies of two-particle interaction vertices. We show results in the weak and strong coupling regime of the half-filled Hubbard model in two dimensions,
illustrating that the method converges quickly where dynamical mean field theory is a good approximation, and show that corrections are large in the strong
correlation regime at intermediate interaction. The fast convergence of dual corrections to dynamical mean field results illustrates the
power of the approach and opens a practical avenue towards the systematic inclusion of non-local correlations in correlated materials simulations. An analysis
of the frequency scale shows that only low-frequency propagators contribute substantially to the diagrams, putting the inclusion of higher order vertices within
reach.\end{abstract}


\maketitle
\section{Introduction}
The dynamical mean field theory,\cite{Metzner89,Georges92,Jarrell92,Georges96} a cornerstone of modern materials simulation,\cite{Kotliar06} has been designed
as an exact method for lattice models in infinite coordination number limits, and has proven to be a useful approximation for the simulation of realistic and
model systems in three and two dimensions. It is based on the realization that, if correlations are purely local, the diagrammatics of an intractable extended
lattice model can be simplified to that of an auxiliary impurity model coupled to a self-consistently adjusted bath. Numerous numerical techniques for the
solution of impurity models exist.\cite{Caffarel94,Bulla08,Gull11_RMP,Zgid12} In the finite coordination number limit relevant for materials simulations, DMFT
provides accurate predictions in the weak and strong coupling limits as well as at high temperature, and qualitatively captures many of the salient features of
correlated electron systems.

Away from these limits, the assumption of local correlations is an approximation. 
The dual fermion method,\cite{Rubtsov08} one of several methods that re-introduce non-local correlations in a systematic
way,\cite{Toschi07,Rubtsov08,Slezak09,Kusunose06,Pollet11,Rubtsov2012,Rohringer2013,Wentzell2015,Li2015,Ayral2015} is based on a Hubbard-Stratonovich transform
of auxiliary `dual' fermionic degrees of freedom chosen such that the difference between the exact single-particle self-energy and its DMFT approximation is
quantified as a perturbation series of impurity quantities in `dual' space.
This dual series is expected to converge quickly where DMFT is accurate. As temperature, interaction, and particle number are changed into regimes in which
strongly non-local correlations are expected, the quality of DMFT as a starting point is less certain and corrections to it are expected to be large, so that
the series may even diverge.

The dual series consists of an infinite number of terms of two-, three-, and higher order impurity vertex diagrams connected by dual propagators. Calculations of this series  have so far been restricted to dual two-particle vertices and to second-order \cite{Rubtsov2009,Brener2008,Antipov2011} or infinite RPA-like ladder diagram approximations.\cite{Hafermann2009, Antipov14,Li2014,Otsuki2014,Otsuki2015,Hirschmeier2015} This restriction stems from two limitations: First, the higher-order impurity vertex functions are difficult to compute numerically. 
Second, the systematic analytic evaluation of dual diagrams with general topology in the absence of a small parameter has proven to be difficult in practice.

As \textcite{Yang11} pointed out, the truncation of the series to two-particle vertices is justified in the context of cluster extensions to the dynamical mean
field theory, where the inverse cluster size $1/N_c$ fulfills the role of a control parameter and the truncated series accelerates convergence from $(1/N_c)^2$
to $(1/N_c)^4$ in two dimensions. It can also be justified in the Falicov-Kimball model at half filling, where any higher order correlation function
vanishes.\cite{Antipov14,Ribic2016} Similarly, near phase transitions, where a divergence of a certain class of ladder diagrams is expected, ladder summations
have been shown to yield the correct critical exponents.\cite{Antipov14,Hirschmeier2015,Rohringer2011} However, outside of these limits the status of diagram
series neglecting certain classes of summations is uncertain, and results previously obtained in other contexts\cite{Bold10,Gukelberger15} have shown that
contributions from all diagram topologies may be expected outside the weakly correlated regime.

A way to sum all contributions to the series is therefore highly desirable. Diagrammatic Monte Carlo \cite{Prokofev98,VanHoucke10} is a generic technique to stochastically
calculate the value of any Feynman diagrammatic series with diagrams of any topology up to any order.
Its power lies in the fact that -- for a convergent series -- all diagrams are stochastically generated without explicit approximation or truncation, such that
the final result is exact up to Monte Carlo errors. These errors converge to zero with the square root of the inverse of the number of diagrams generated.

In this paper we present an implementation of the stochastic summation of the complete dual series using a diagrammatic Monte Carlo algorithm. We show a quick
convergence at low order in the weak and the strong coupling limits. We also show that, away from the weak coupling limit, non-ladder diagrams contribute
substantially to the series. Finally, we show that due to the rapid convergence of dual propagators in frequency space, only low-frequency terms contribute to
higher order diagrams. For these, the calculation of three- and higher particle number vertices may be technically feasible.

The remainder of this paper is organized as follows: Section \ref{sec:form} introduces the model, the DMFT approximation, and the dual fermion series. Section \ref{sec:stoch} introduces the stochastic sampling procedure. Section \ref{sec:res} discusses results for the half-filled Hubbard model at various interaction strengths, and Section \ref{sec:conc} describes conclusions.
\section{Model and Dual Fermion Series}\label{sec:form}
While the method we present here is applicable to any fermionic lattice model with local interactions, we focus on the example of the Hubbard model in two dimensions for which ample test and benchmark results exist.\cite{LeBlanc15} The Hamiltonian in a mixed real/momentum space notation is
\begin{align}
H = \sum_{k\sigma} \left(\varepsilon_k - \mu\right) c^\dagger_{k\sigma} c_{k\sigma} + U \sum_i n_{i\uparrow}n_{i\downarrow}.\label{Eq:Ham}
\end{align}
$k = \{ k_x, k_y \}$ denotes a vector in the reciprocal space, $\varepsilon_k = -2t (\cos k_x + \cos k_y)$ is the dispersion, $\mu$ is the chemical potential, $i$ labels sites, $\sigma = \uparrow,\downarrow$ spins, and $U$ the on-site Coulomb repulsion. We work in the imaginary time effective action representation of the problem,
\begin{equation}\label{Eq:ActionTau}
S = \int_0^\beta d\tau \left(\sum_{k\sigma} c_{k\sigma}^\dagger (\tau) \frac{d}{d\tau}c_{k\sigma} (\tau) + H(\tau) \right).
\end{equation}


We start our derivation by adding and removing an (initially arbitrary) spin-symmetric local hybridization function $\Delta(i\omega_n)$
\begin{align}\label{Eq:Action}
S & = \sum_{i}S_i^{\mathrm{imp}} - \sum_{\omega k \sigma}(\Delta(i\omega_n) - \varepsilon_k) c^\dagger_{\omega k \sigma} c_{\omega k\sigma} \\
S_i^{\mathrm{imp}} & = \sum_{\omega\sigma} \left( \Delta(i\omega_n) - i\omega_n - \mu\right) c^\dagger_{i\omega \sigma} c_{i\omega \sigma} + \label{Eq:ImpAction}\\ & + U \int_0^\beta d\tau n_{i\downarrow}(\tau) n_{i\uparrow}(\tau). \notag
\end{align}
Eq.~\ref{Eq:Action} redefines the action (\ref{Eq:ActionTau}) as a sum of multiple identical ``impurities'' coupled by the dispersion, and while
Eq.~\ref{Eq:Action} remains intractable, the interacting impurity Green's function $G^{\mathrm{imp}}(\tau)=-\langle c(\tau)c^\dagger(0)\rangle_{\mathrm{imp}}$ or
its Fourier transform $G^{\mathrm{imp}}(i\omega_n)$  can be obtained for any given $\Delta$. \cite{Werner06,Gull11_RMP,Gull11_ALPSDMFT,ALPS20} We assume
paramagnetic spin symmetry and omit explicit spin indices unless needed.

The dual fermion method, introduced by \textcite{Rubtsov08}, performs a perturbative expansion of Eq.~\ref{Eq:Action} based on the solution of an
impurity problem with an (initially arbitrary) hybridization $\Delta$. The method is exact in the sense that if all terms of the expansion are summed, the exact
solution of Eq.~(\ref{Eq:Ham}) is recovered. 
While the choice of $\Delta$ is arbitrary, it is most convenient to use the result of a self-consistent solution
of a DMFT problem. In that case, the impurity propagator already contains all local correlations. We will consider only this case in the rest of this paper.

The dual fermion method proceeds from Eq.~\ref{Eq:Action} by introducing an auxiliary set of degrees of freedom, the `dual' fermions $\xi$, through a
Hubbard-Stratonovich transformation:
\begin{multline}\label{Eq:MixedAction}
S = \sum_{i \omega \sigma} \left[ S^{\mathrm{imp}}_i + [G^{\mathrm{imp}}(i\omega_n)]^{-1} \left( c^{\dagger}_{i\omega\sigma}\xi_{i\omega\sigma} + \xi^\dagger_{i\omega\sigma} c_{i \omega\sigma} \right)\right] \\ + \sum_{\omega k \sigma} [G^{\mathrm{imp}}(i\omega_n)]^{-2} (\Delta(i\omega_n) - \varepsilon_k)^{-1} \xi^\dagger_{\omega k \sigma} \xi_{\omega k \sigma} 
\end{multline}
These dual fermions decouple the interacting impurities in Eq.~(\ref{Eq:Action}). All $c$-electron terms in  Eq.~(\ref{Eq:MixedAction}) are factorized for every
lattice site and can be integrated out in a cumulant expansion of $S^{\mathrm{imp}}$, resulting in an effective action for $\xi$ fermions only:
\begin{equation}\label{eq:dual_action}
S = -\sum_{\omega k \sigma} \left[\tilde G(i\omega; k)\right] ^ {-1} \xi^\dagger_{\omega k \sigma} \xi_{\omega k \sigma} + \sum_{i} V_{i}[\xi^\dagger, \xi],
\end{equation}
where $\tilde G (i\omega, k) = G^{\mathrm{DMFT}}(i\omega_n, k) - G^{\mathrm{imp}}(i\omega_n)$ is the bare propagator of dual fermions\cite{Otsuki2014}, and the
dual interactions are defined as
\begin{align}
V_i = -\sum_{n=2}^{\infty} (-1)^{n}\frac{\gamma^{(2n)}_{i_1\hdots {i_{2n}}}}{(n!)^2} \prod_{m=1}^{n} (\xi^\dagger_{i_m}\xi_{i_{m+1}})^{n},
\end{align}
with $i_m$ labeling the combination of site index, Matsubara frequency and spin, $i_m = \{i; \omega_m; \sigma_m\}$, and $\gamma^{(2n)}$ is the one-particle
reducible impurity vertex.\cite{Rohringer12} The lowest order, $n=2$, is
\begin{multline}
\gamma^{(4)}_{i_1i_2i_3i_4} = [G^\mathrm{imp}_{i_1}]^{-1}[G^\mathrm{imp}_{i_2}]^{-1}[G^\mathrm{imp}_{i_3}]^{-1}[G^\mathrm{imp}_{i_4}]^{-1} \times \\   \left( \langle c_{i_1} c^\dagger_{i_2} c_{i_3} c^\dagger_{i_4} \rangle_{\mathrm{imp}} -  G_{i_1}^\mathrm{imp} G_{i_3}^\mathrm{imp} \delta_{12} + G^\mathrm{imp}_{i_1} G^\mathrm{imp}_{i_2} \delta_{13} \right)
\end{multline}
Higher order vertices are obtained by subtracting all reducible combinations from the  $n$-particle impurity propagator. 

Eq.~\ref{eq:dual_action} is an exact reformulation of Eq.~\ref{Eq:Action}, and all dual correlators have exact lattice counterparts. In particular, the
single-particle lattice self-energy is
\begin{equation}\label{eq:sigma_dual}
\Sigma (i\omega_n, k) = \Sigma^{\mathrm{imp}} (i\omega_n) + \frac{\tilde \Sigma (i\omega, k)}{1 + G^{\mathrm{imp}}(i\omega)\tilde\Sigma(i\omega, k)},
\end{equation}
where $\tilde \Sigma(i\omega, k)$ is the self-energy of the dual fermions.\cite{Hafermann08}

The primary object we consider in this paper is the dual Luttinger-Ward functional $\tilde \Phi$ which satisfies $\tilde \Sigma=\frac{\delta\tilde\Phi}{\delta
\tilde G}$. It is obtained through a perturbative expansion of $S$ in the dual interaction $V$ as a sum over all two-particle irreducible diagrams
\cite{Baym1961} to the partition function (\ref{eq:dual_action}):
\begin{equation}
\tilde \Phi= \sum_n \frac{1}{n!} \langle \prod_{i=1}^{n} V_i \rangle_{\mathrm{connected}},
\label{eq:DualPhi_Generic}\end{equation}
where averaging is performed over the noninteracting part of the dual fermion action $S_0 = \sum_{\omega k \sigma} \left[ - \tilde G (i\omega_n, k) \right]^{-1} \xi^\dagger_{\omega k \sigma} \xi_{\omega k \sigma}$.

The dual fermion expansion around the dynamical mean field theory has remarkable properties. First, as DMFT obtains the correct local physics, any local dual contribution is zero. This implies that both the Hartree and the Fock diagram of the series (as well as any diagram containing Hartree or Fock contributions) vanish, and substantially reduces the number of diagrams.
Second, as both $G^\text{DMFT}$ and $G^\text{imp}$ have a high frequency behavior $\sim \frac{1}{i\omega_n}$, the frequency dependence of $\tilde G$ is $\sim \frac{1}{i\omega_n^2}$. 
As a consequence, high order diagrams with many propagators are mostly confined to frequencies near zero, further reducing the size of diagram space.

Practical implementations of the method typically make two approximations. First, the complete solution of Eq.~(\ref{eq:DualPhi_Generic}) requires the
computation of an infinite series of $2n$-operator impurity vertex function $\gamma^{(2n)}$, which is typically truncated to two-particle vertices. Second, the expansion requires a summation over all possible diagrams, and so far always has either been truncated at second order or approximated by an
RPA-like `ladder' summation over vertex diagrams. 

\begin{figure}[tbh]
\includegraphics[width=0.32\columnwidth]{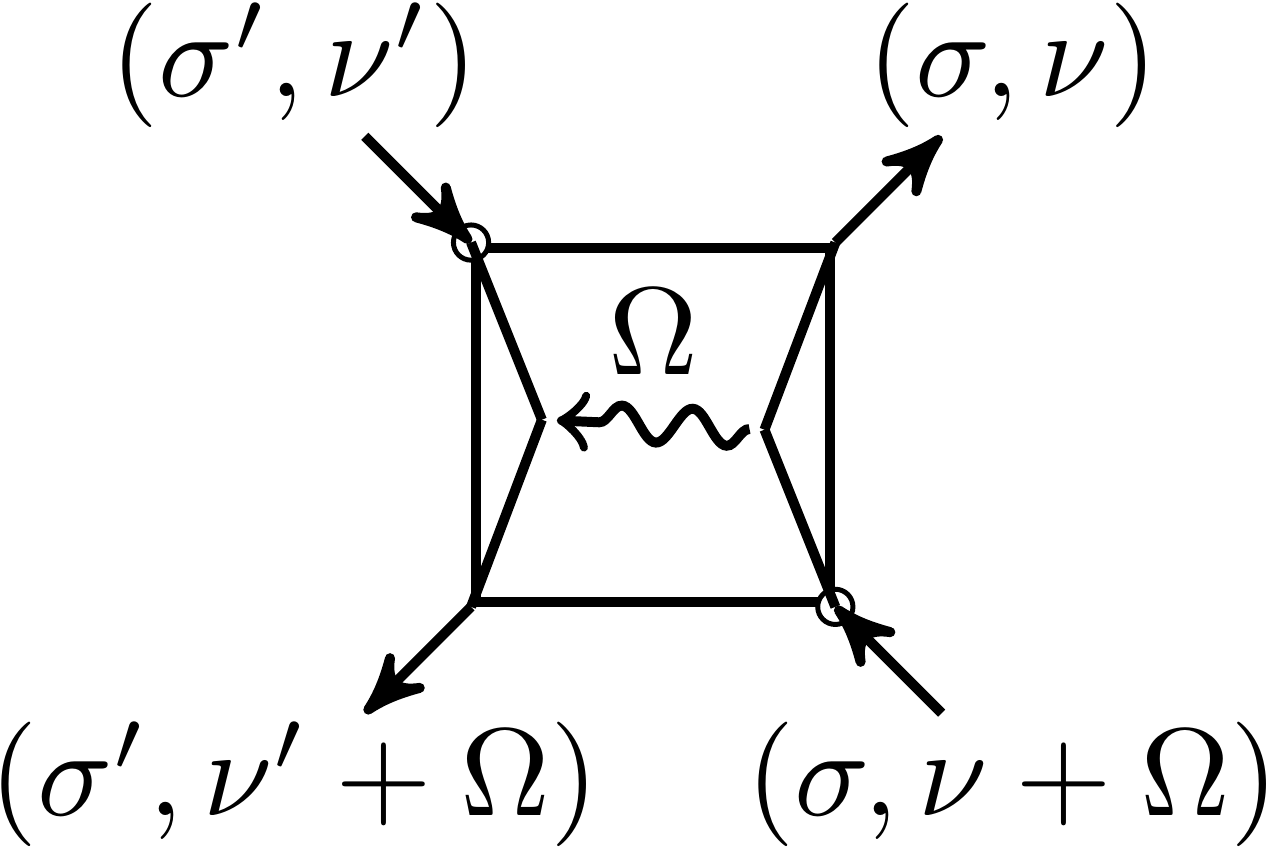}
\includegraphics[width=0.52\columnwidth]{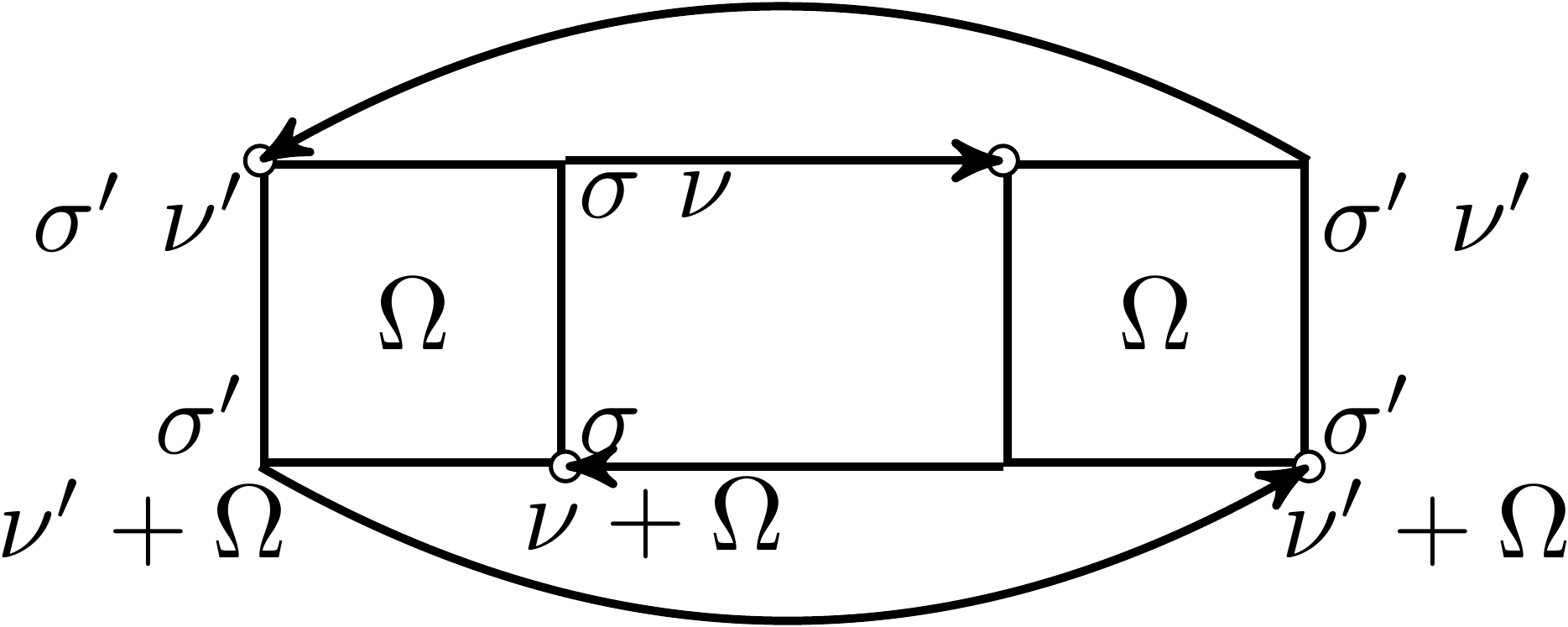}
\caption{Left panel: Diagram notation for a dual fermion vertex. Two incoming lines with momenta/energies $\nu+\Omega$ and $\nu'$ and spin $\sigma$ and $\sigma'$ scatter into two outgoing lines with momenta $\nu'+\Omega$ and $\nu$. A momentum $\Omega$ is transferred from right to left. Right panel: lowest order diagram. For clarity we omit the arrow inside the vertex.} \label{fig:Vertex}
\end{figure}

\section{Stochastic Sampling of the Dual Fermion Series}\label{sec:stoch}

In this work, we instead consider the full series of all dual diagrams with two-particle vertices using a stochastic diagrammatic Monte Carlo sampling
procedure. We neglect three-particle and higher order impurity vertices. The diagram series is given by Eq.~\ref{eq:DualPhi_Generic} and reads
\begin{multline}
\tilde{\Phi} = \sum_n \frac{1}{n!} \int_{C} \sum_{i^{(1)}=1}^{n}\sum_{i^{(2)}=1}^{n}\ldots\sum_{i^{(n)}=1}^{n}\prod_{k=1}^{n} \\
          (-\frac{1}{4}\gamma_{i^{(k)}_{1234}} \xi_{i^{(k)}_1}\xi^{\dagger}_{i^{(k)}_2}\xi_{i^{(k)}_3}\xi^{\dagger}_{i^{(k)}_4}) e^{-S^{d}_{0}\left[\xi^\dagger,\xi\right]}
          D[\xi,\xi^{*}] \\
          = \sum_{n} \sum_{i^{(1)}=1}^{n}\sum_{i^{(2)}=1}^{n}\ldots\sum_{i^{(n)}=1}^{n} \mathcal{W}_{n}(\mathcal{C}),
\end{multline}
where the $i^{(m)} = (\nu_m,\nu_m',\sigma_m,\sigma_m')$ is a composite index for the $m$-th vertex, $\nu = (\omega, k)$, and $\mathcal{W}_{n}(\mathcal{C})$ is
the contribution of the each term to the dual Luttinger-Ward functional: a Feynman diagram. These diagrams build the Monte Carlo configurations for our
stochastic sampling.
Each diagram evaluates to a (complex) number
\begin{multline}
\mathcal{W}_{n}(\mathcal{C}) = \frac{(-1)^n}{n!}\left(\frac{1}{4}\right)^{n} \prod_{m=1}^{n} \gamma^{\sigma_m\sigma'_m}_{\nu_m\nu'_m\Omega_m}
\tilde{G}_{\sigma_m}(\nu_m) \times \\ \tilde{G}_{\sigma'_m}(\nu'_m+\Omega_m)
\end{multline}
and is uniquely defined by a graph: a set of interaction vertices connected by propagators, defined as $\langle\xi_{i_m}\xi^{\dagger}_{j_m}\rangle_{S_0} =
-\tilde{G}_{\sigma_m}(\nu_m)$ and appropriately anti-symmetrized vertex functions $\gamma^{\sigma_m\sigma'_m}_{\nu_m\nu'_m\Omega_m} = \gamma_{i^{m}}$. Momentum
and energy are conserved at each vertex, and each vertex consists of a scattering process of two dual fermions with momentum transfer $\Omega$. The vertices are
graphically illustrated in the left panel of Fig. \ref{fig:Vertex}: A dual particle with momentum end energy $\nu+\Omega$ and spin $\sigma$ scatters with a
particle with momentum and energy $\nu'$ and spin $\sigma'$ and imparts on it a momentum/energy of $\Omega$.
Vertices carry a weight $\frac{1}{4}\gamma_{\nu\nu'\Omega}^{\sigma\sigma'}$, each line carries spin and momentum/energy 
and evaluates to $\tilde{G}_{\sigma_m}(\nu_m)$. A typical second order diagram is illustrated in the right panel of Fig.~\ref{fig:Vertex}, where two vertices
with identical momentum transfer and two fermion loops are present.

\subsubsection{Monte Carlo sampling procedure}
\begin{figure}[tbh]
\includegraphics[width=0.45\columnwidth]{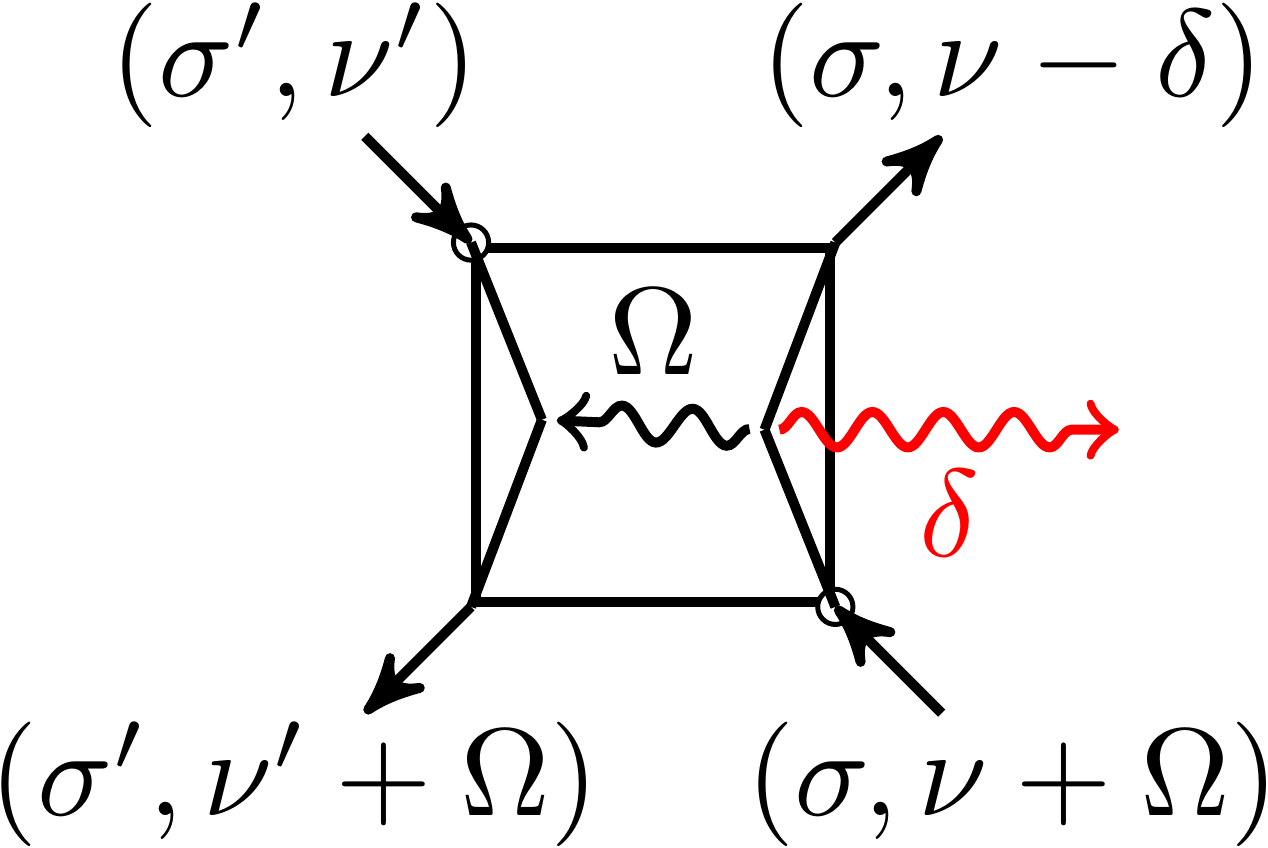}
\includegraphics[width=0.45\columnwidth]{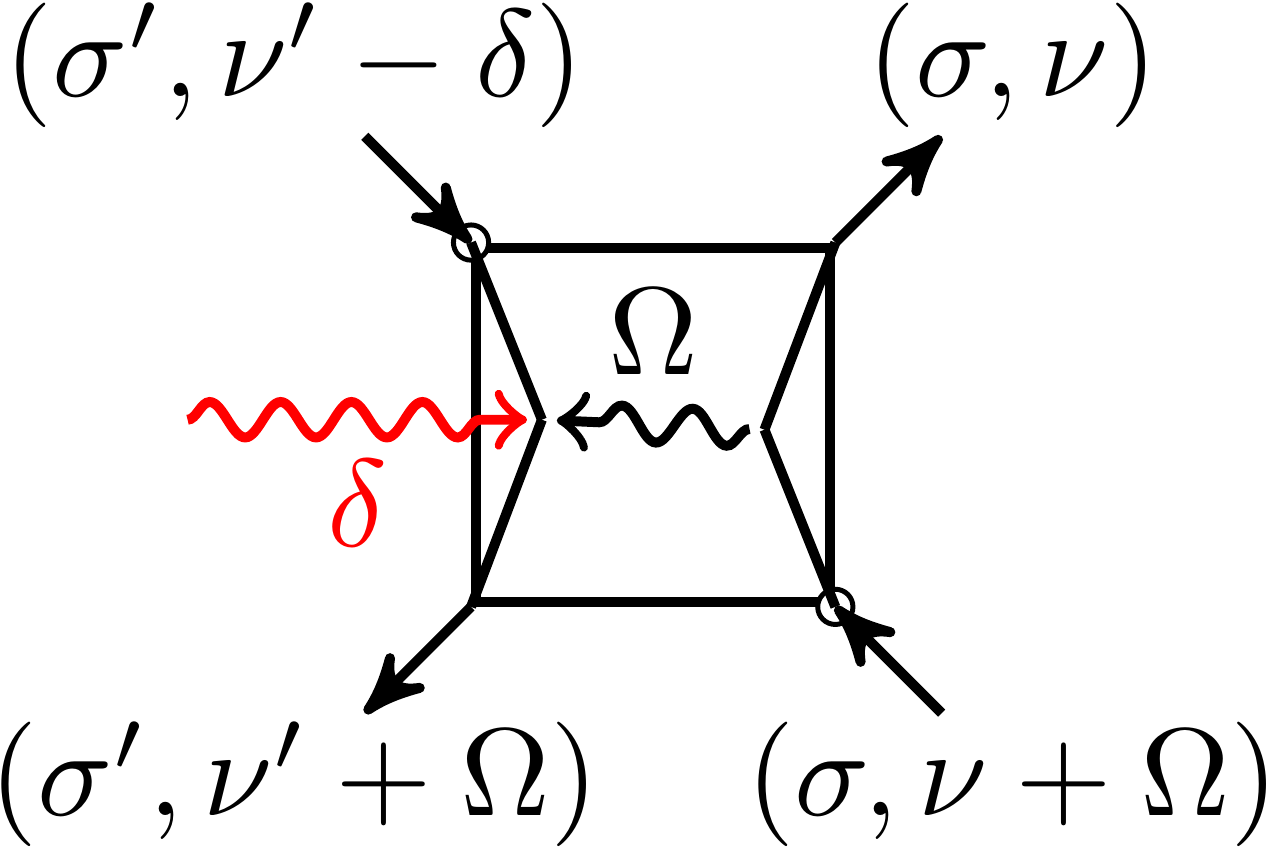}
\caption{Auxiliary `worm' diagrams. Left panel: a worm tail is attached to the spin-$\sigma$ line and  carries away momentum $\delta$. Right panel: a worm head, attached to the spin $\sigma'$ line, carries in momentum $\delta$ (right). Note that momentum is conserved at each vertex and head or tails can be attached to either line.} \label{fig:VertexWorm}
\end{figure}

To stochastically sample all diagrams, we define a set of updates to change diagram order, diagram topology, the energy and momentum of propagators and vertices,
as well as the spin of propagator lines, and perform a Markov chain Monte Carlo random walk in diagram space. Our method is an adaptation of the established diagrammatic Monte Carlo algorithms\cite{Prokofev98,Kozik10,VanHoucke10} but is formulated in Matsubara frequency rather than imaginary time space. To ensure that all possible diagrams are generated
(ergodicity), while respecting spin, energy, and momentum conservation, we found it necessary to enlarge our space of diagrams and introduce
``worms'':\cite{Prokofev98} auxiliary bosonic lines (of which we have at most one at any time) that carry momentum and frequency from one part of the diagram to
another. These `worms' are attached to two possible positions of our vertices, see Fig.~\ref{fig:VertexWorm} for an illustration.
As the weight of a diagram in worm space we choose the product of propagator and  vertex factors that we have in Luttinger-Ward diagram space, multiplied by a
parameter $\eta$ which is chosen such that about equal time is spent in worm and Luttinger-Ward diagram space. A sequence of updates typically proceeds from a
diagram, inserting a worm line with a random momentum and energy, using it to change topology or diagram order, and finally removing it. Unlike in other
algorithms where worms represent Green's function configurations, worm configurations have no physical meaning and are purely used as an auxiliary formulation
to update diagrams. We found the following set of updates necessary to correctly generate all topologies: worm insertion and removal, worm move, change of
topology, change of order, and spin-flip updates.

\begin{figure}[tbh]
\includegraphics[width=0.45\columnwidth]{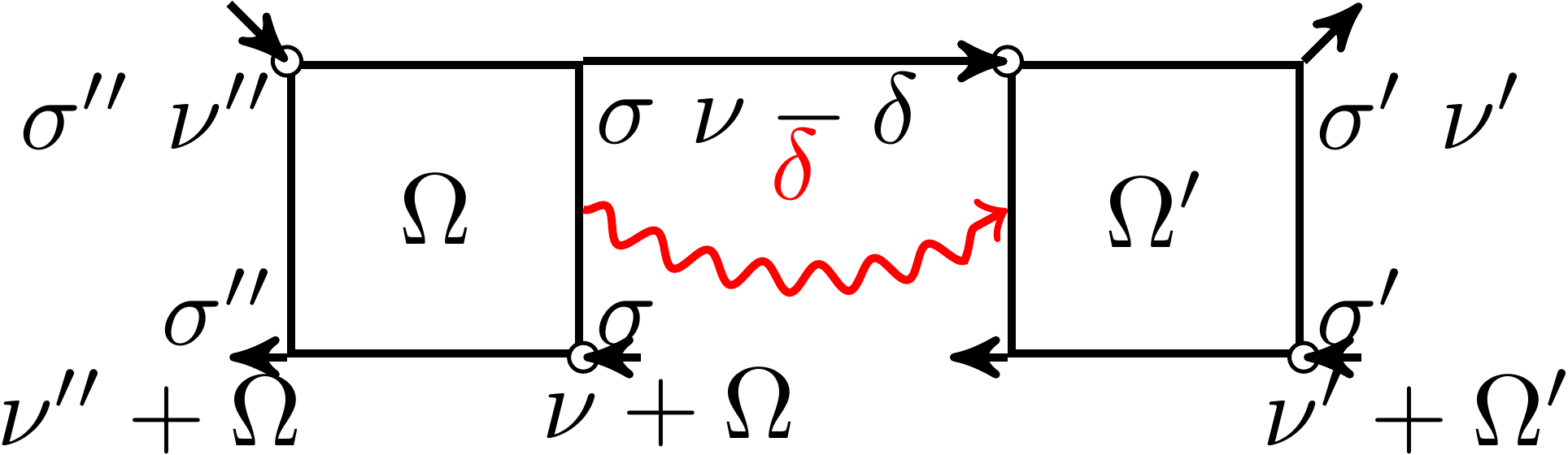}
\includegraphics[width=0.22\columnwidth]{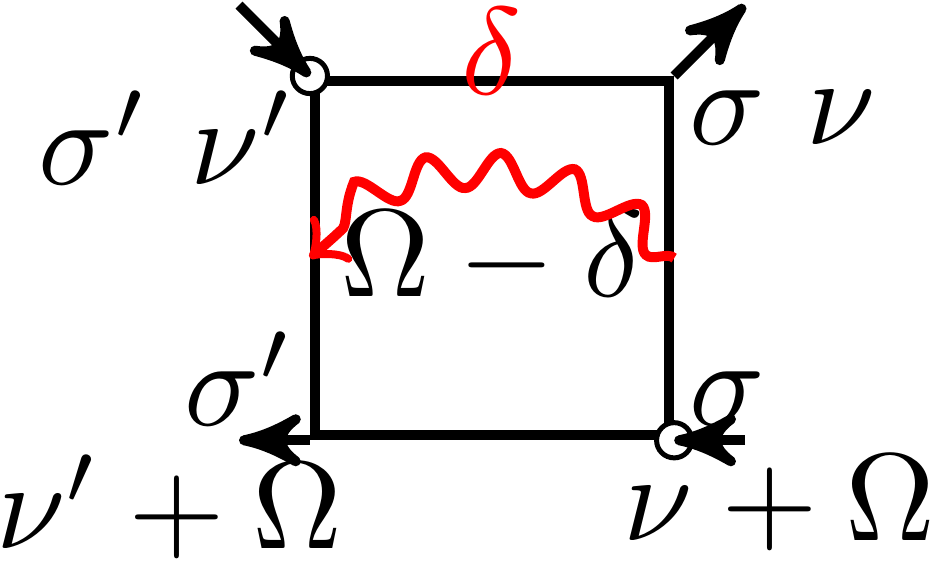}
\caption{Two complementary ways of inserting an auxiliary `worm' diagram line carrying momentum $\delta$ into a second order Luttinger-Ward diagram.}
\label{fig:WormInsertion}
\end{figure}
\subsubsection{Insertion and Removal of a Worm}
We start the discussion of the Monte Carlo updates with the insertion and removal of an auxiliary boson `worm' line. This update transitions between
Luttinger-Ward and worm space. For inserting a worm, we choose a random propagator and a random worm energy $\delta$, modify the propagator energy $\nu$ to
$\nu-\delta$, attach the worm tail to the source of the propagator and the worm head to the target of the propagator. The resulting diagram conserves momentum
and is illustrated in the left panel of Fig.~\ref{fig:WormInsertion}. The reverse move takes an existing worm and proposes to remove it if it goes along a
propagator, while changing the propagator energy from $\nu-\delta$ to $\nu$.

A second, complementary update inserts a worm with energy $\delta$ on a vertex or removes it from a vertex. In this case, the vertex frequency is modified from
$\Omega$ to $\Omega-\delta$ (insertion) or $\Omega-\delta$ to $\Omega$ (removal). These updates are illustrated in the right panel of
Fig.~\ref{fig:WormInsertion}. Satisfying detailed balance requires taking into account the proper update proposal probabilities. We use Metropolis updates, for
both updates the acceptance criteria are the same and given by
\begin{align}
p^\text{worm insert}(\delta)=\min\left(1,\frac{p_{new}}{p_{old}} \frac{k}{p(\delta)} \eta \right) \label{eq:worm_insert}
\\
p^\text{worm remove}(\delta)=\min\left(1,\frac{p_{new}}{p_{old}} \frac{p(\delta)}{k} \frac{1}{\eta}\right),
\label{eq:worm_remove}
\end{align}
where $k$ denotes the expansion order and $p(\delta)$ the probability for proposing a random bosonic frequency $\delta$.

\begin{figure}[tbh]
\includegraphics[width=0.45\columnwidth]{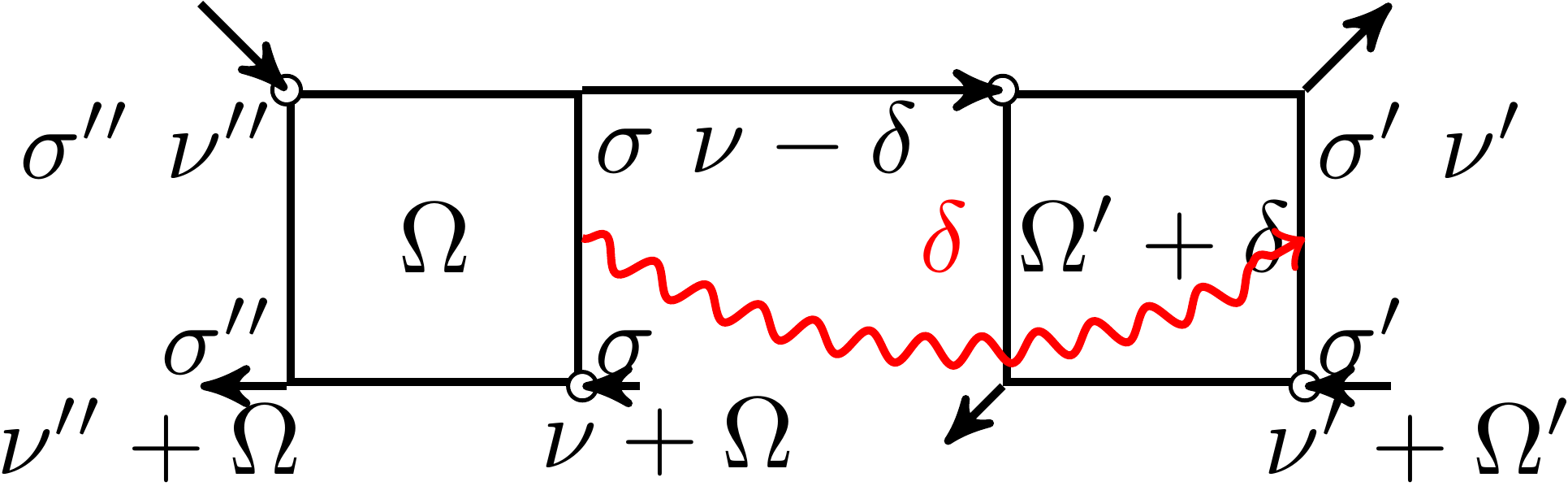}
\includegraphics[width=0.45\columnwidth]{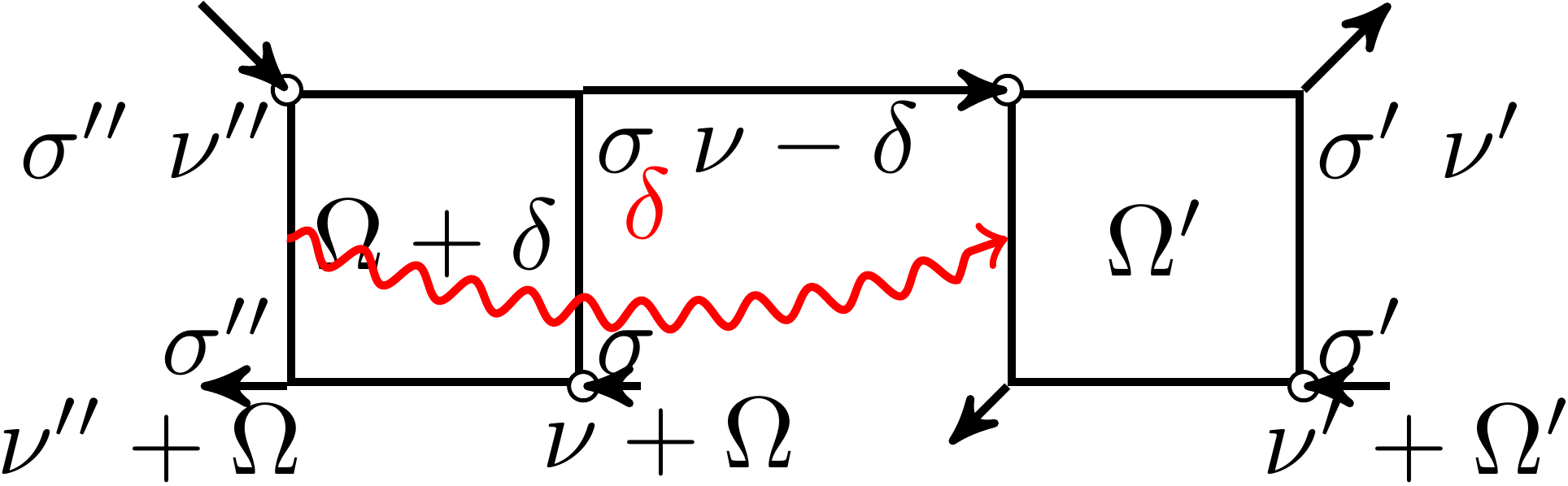}
\includegraphics[width=0.65\columnwidth]{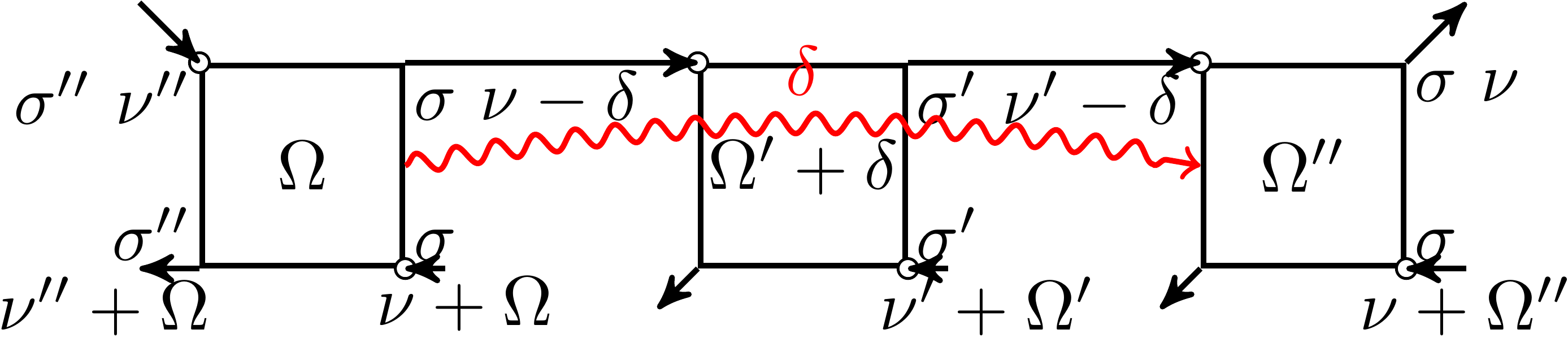}
\caption{`worm move' updates: starting from a worm configuration as in the left panel of Fig.~\ref{fig:WormInsertion}, the worm head is (top left) moved across a vertex, changing its momentum transfer from $\Omega$ to $\Omega-\delta$, or (top right) its tail is moved across a vertex, changing its momentum from $\Omega$ to $\Omega +\delta$. Bottom panel: a worm head is moved along a propagator line, changing its momentum from $\nu'$ to $\nu'-\delta$.} \label{fig:WormMove}
\end{figure}

\subsubsection{Move of a worm}
The worm `move' update moves a worm head or tail along parts of the diagram. Several distinct possibilities exist: First, a worm `head' is moved forward along
an outgoing propagator line, in which case the value of the propagator is changed from $\nu$ to $\nu+\delta$. Second, a worm head is moved backward along an
outgoing vertex line, in this case the value of the propagator is changed from $\nu$ to $\nu - \delta$. These two updates balance each other. Third, a worm tail
is moved along an outgoing propagator line. In this case, the propagator's energy is changed from $\nu$ to $\nu - \delta$. Fourth, a worm tail is moved backward
along an incoming propagator line. In this case, the propagator's energy is changed from $\nu$ to $\nu+\delta$. These updates, too, balance each other. Fifth, a
worm head is moved from the right side of a vertex to the left side of a vertex, in which case the momentum transfer of that vertex is changed from $\Omega$ to
$\Omega -\delta$. Sixth, a worm head is moved from the left side of a vertex to the right side of a vertex, in which case the momentum transfer of the vertex is
changed from $\Omega$ to $\Omega + \delta$. Similar relations for the move of a worm tail across a vertex and for the move of a worm head or  tail forward or
backward along an incoming propagator line are easily derived. Fig.~\ref{fig:WormMove} illustrates some of these updates. As the proposal probability of a move and its inverse move are identical, no factors appear in the Metropolis criterion.


\begin{figure}[tbh]
\includegraphics[width=0.85\columnwidth]{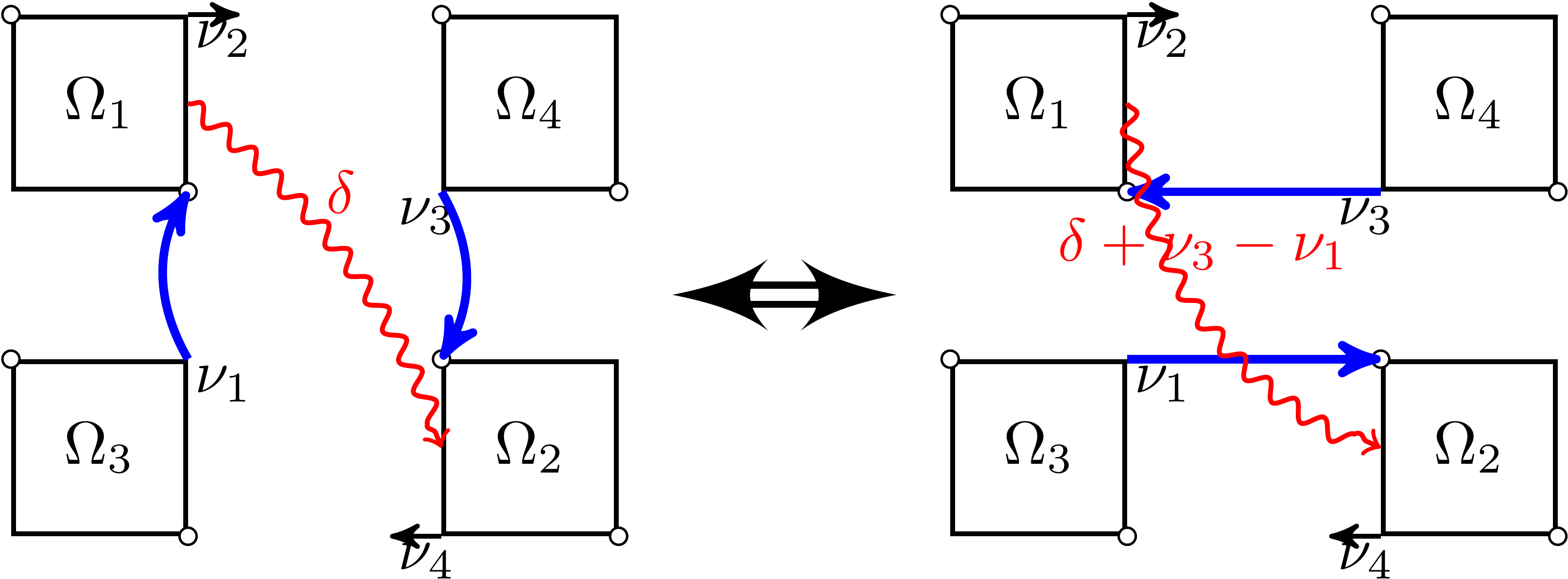}
\caption{Updates changing diagram topology: A worm connecting two diagrams reconnects their outgoing vertex lines.}\label{fig:WormReconnect}
\end{figure}

\subsubsection{Change of topology}
Changes of diagram topology are best done in worm space. To set the stage we assume an initial configuration in which a worm connects vertex $1$ to vertex $2$,
with the incoming propagator (energy $\nu_1$) of vertex $1$ on the side of the worm  pointing from vertex $3$ and the incoming propagator on vertex $2$ (energy
$\nu_3$) pointing from vertex $4$. This configuration is shown in the left panel of Fig.~\ref{fig:WormReconnect}. It is then possible to design an update that
reconnects the propagator of vertex $4$ to vertex $1$ and of vertex $3$ to vertex $2$, as illustrated in the right panel of Fig.~\ref{fig:WormReconnect}.
Without the presence of a worm, such an update would most likely violate momentum conservation. Here, momentum conservation can be restored by changing the worm
momentum from $\delta$ to $\delta+\nu_3-\nu_1$. This update is its own reverse update.

\begin{figure}
\includegraphics[width=0.95\columnwidth]{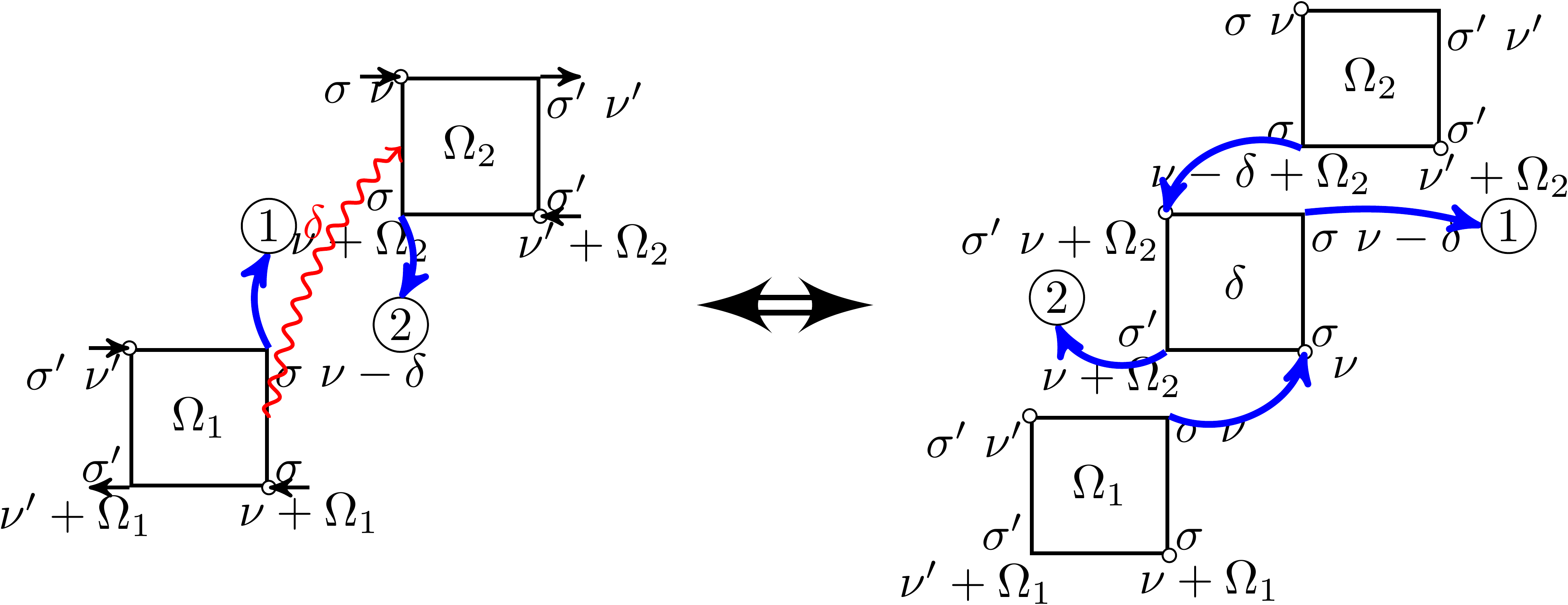}
\caption{Updates changing diagram expansion order: As read from left to right: update changing a worm space configuration with $n$ vertices into a
Luttinger-Ward diagram configuration with $n+1$ vertices, by inserting a new vertex with momentum transfer $\delta$. As read from right to left: update changing
a Luttinger-Ward diagram configuration with $n+1$ vertices into a worm configuration with $n$ vertices by removing a vertex.}\label{fig:SpawnDissolve}
\end{figure}

\subsubsection{Change of diagram order}
In order to change the diagram order of the expansion, we employ an update that takes a `worm' diagram and  inserts in its place a new vertex or, conversely,
takes a Luttinger-Ward diagram configuration and replaces one of its vertices by a worm line. In this update, illustrated in Fig.~\ref{fig:SpawnDissolve}, the
worm momentum $\delta$ is converted into the vertex transfer momentum. Starting from a worm configuration connecting two vertices $1$ and $2$, the update
consists of three steps. First, a new vertex with momentum transfer $\delta$ is created and its outgoing legs are connected to the targets of the outgoing legs
of vertices $1$ and $2$. Second, the incoming legs of the new vertex are connected to the sources of the outgoing legs of vertices $1$ and $2$. Third, the
momenta of the newly created legs are adjusted such that momentum conservation at each vertex is satisfied. The reverse update consists of selecting a vertex,
replacing it by a worm line, and adjusting momenta such that momentum conservation is respected, as illustrated in the right panel of Fig.~\ref{fig:SpawnDissolve}.


\begin{figure}[tbh]
\includegraphics[width=0.5\columnwidth]{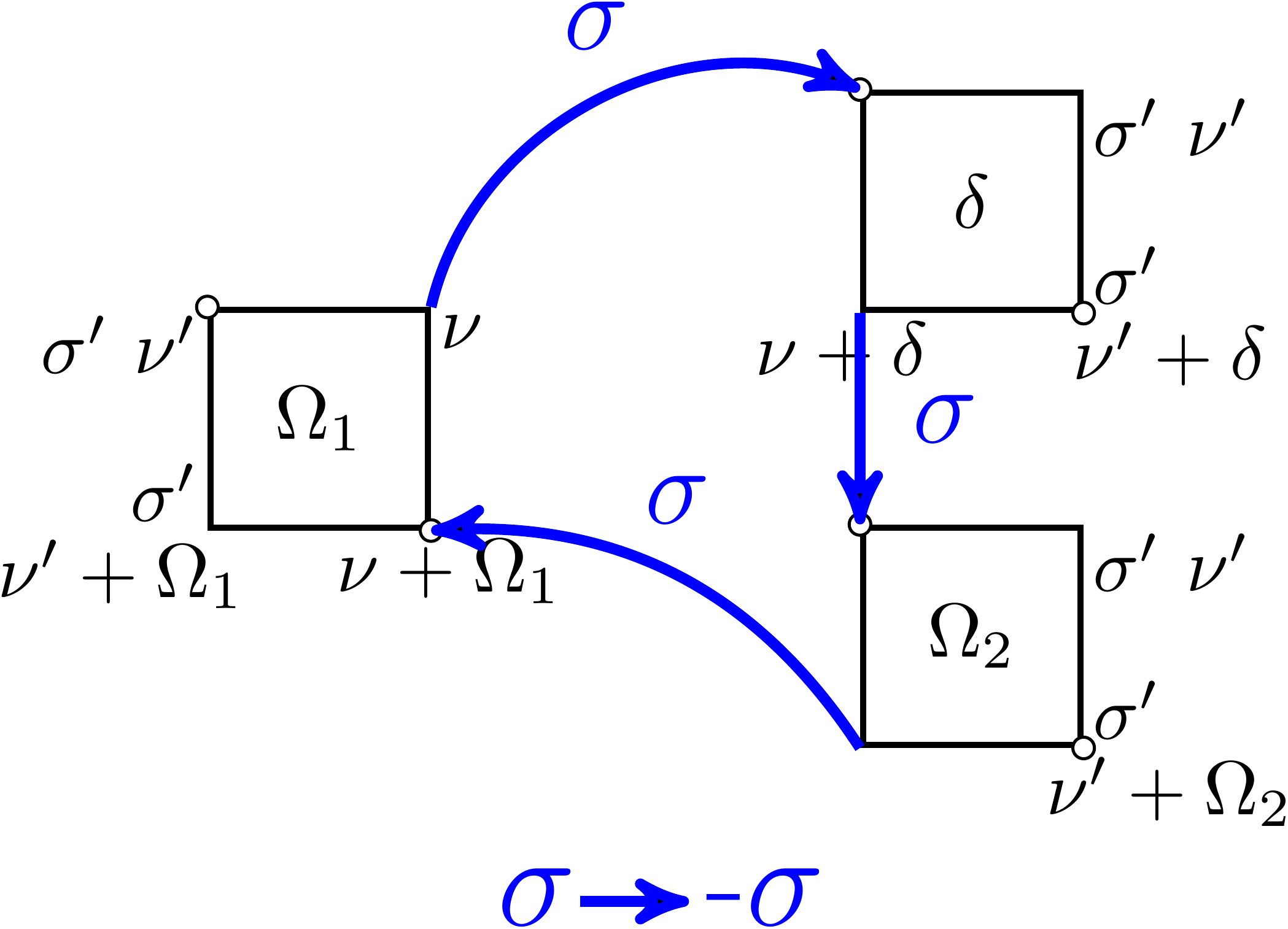}
\caption{Example of a fermion loop identified by randomly selecting a propagator and following one of its fermion lines until it reaches the original vertex. The `spin flip' update then proceeds by flipping the spin of all propagators and vertices along this loop.}\label{fig:Noqloop}
\end{figure}
\subsubsection{Spin-flip update}
Since new vertices always inherit the spins of the propagators from which they have been created, an additional update is necessary: the change of a fermion
loop from spin $\sigma$ to spin $-\sigma$. For this purpose we identify a fermion loop by following the diagram along a propagator, without traversing a vertex,
until the loop closes. Such a loop is non-local as it covers at least two vertices (Hartree and Fock diagrams are zero) but may cover a large fraction of the diagram vertices. We then propose to flip the spin of that propagator from $\sigma$ to $-\sigma$ and accept or reject according to the
Metropolis criterion. Fig.~\ref{fig:Noqloop} illustrates such a loop. This update is self-balanced.

\begin{figure}[tbh]
\includegraphics[width=0.95\columnwidth]{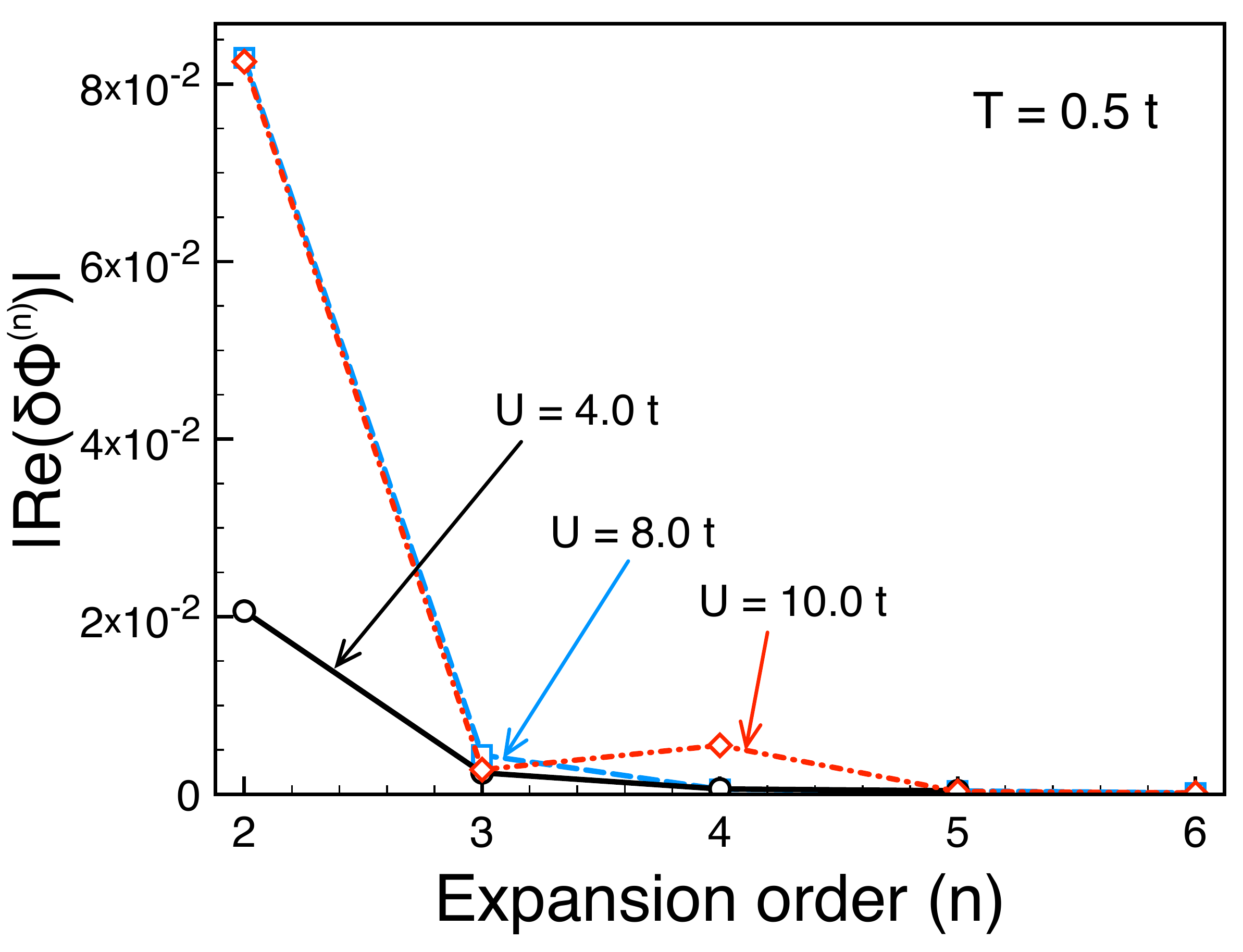}
\caption{Contribution to the real part of the Luttinger-Ward functional of the dual fermion expansion as a function of diagram expansion order.
Hubbard model, half filling, $T/t=0.5$ and $U/t=4.0, 8.0, 10.0$.}\label{fig:Orders}
\end{figure}
\section{Results}\label{sec:res}
In this section we present results obtained by the diagrammatic Monte Carlo dual fermion algorithm. We first illustrate the most important technical aspects of
the diagrammatic sampling, then introduce results for the main quantity obtained in the algorithm: the dual self-energy, and finally present results for the
physical self-energy and a comparison to results from another method. We defer a detailed analysis of the physics of the half-filled and doped two-dimensional
Hubbard model to a later publication.
 
We start our discussion of the results with a plot for the contribution of diagrams at each order to the dual Luttinger-Ward functional for different
interaction strength, Fig.~\ref{fig:Orders}, at a relatively high temperature $T/t=0.5$. At intermediate to weak coupling ($U/t=4$, black circles), second order corrections capture almost the entire
difference to the DMFT result. Contributions of terms with more than three vertices are negligibly small.
At an interaction strength of the bandwidth ($U/t=8$, blue squares), we find that  second order contributions are still dominant, and the magnitude of their contribution has increased
substantially. Fourth and higher order contributions are essentially zero.
At an interaction strength larger than the bandwidth, $U/t=10$, deviations from diagrams containing more than three interaction vertices are visible, but the
series is still convergent at an expansion order of five. Varying doping, temperature, and other parameters can push the series to
regions where much higher order diagrams are important or where it eventually diverges.

\begin{figure}[bth]
\includegraphics[width=0.85\columnwidth]{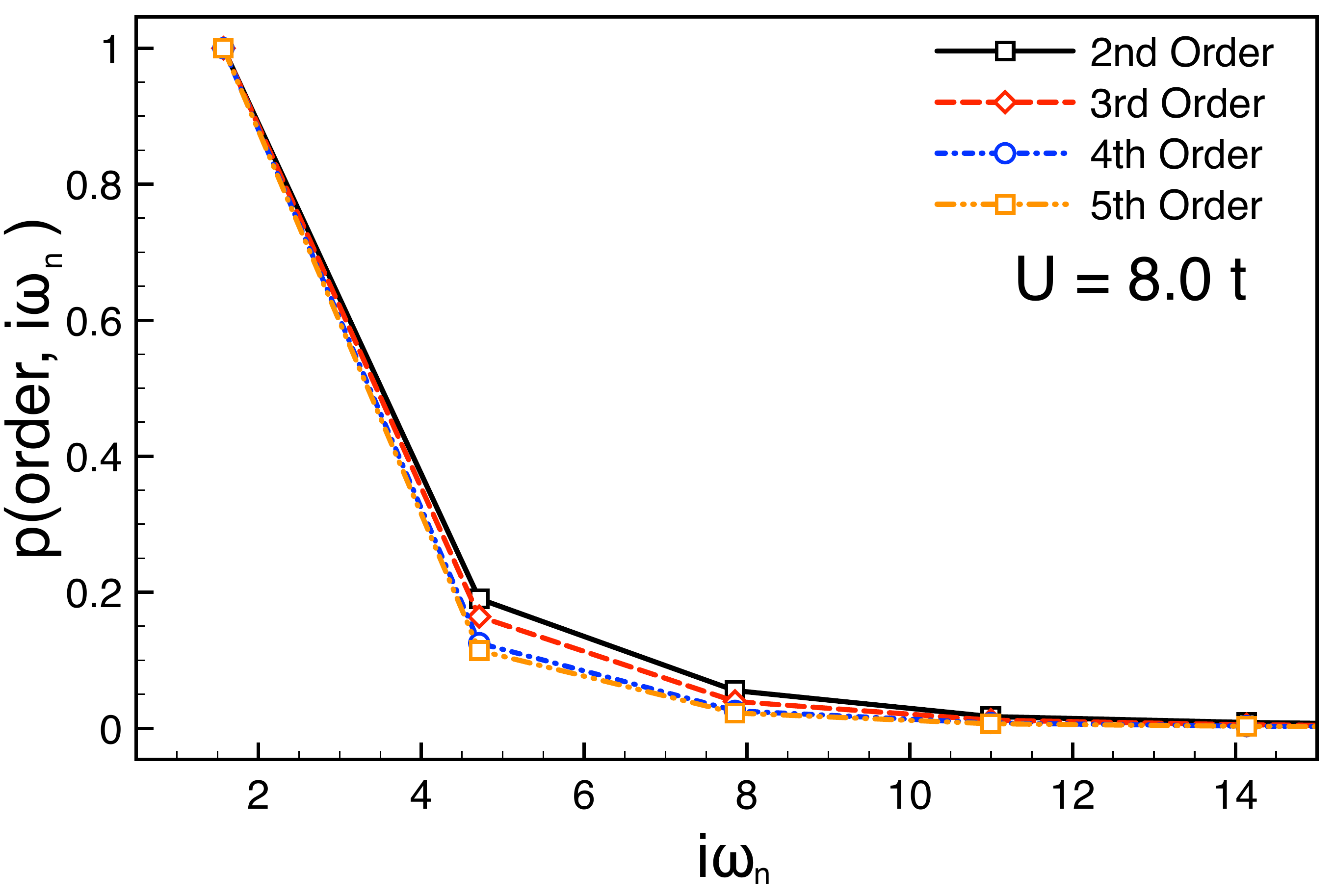}
\caption{Probability that a dual diagram at order $2$, $3$, $4$, or $5$ contains propagators with frequency $i\omega_n$, normalized to the probability of a diagram at the same order containing a propagator at the lowest Matsubara frequency. Data for $T/t=0.5$ and $U/t=8.0$.}
\label{fig:OrderMatsubara}
\end{figure}

The Monte Carlo random walk automatically generates diagrams of the dual Luttinger Ward functional with the weight that they contribute to $\tilde \Phi$. Fig.~\ref{fig:OrderMatsubara} shows the distribution of propagator lines contained in the diagrams generated by the Monte Carlo random walk as a function of frequency, resolved by expansion order. As contributions from higher expansion orders are strongly suppressed (see Fig.~\ref{fig:Orders}), we normalize data for each order to the contribution of the lowest Matsubara frequency at that order. 

It is apparent that only contributions from the lowest few Matsubara frequencies are generated, implying that contributions containing high frequency diagrams are strongly suppressed.

This strong suppression of higher  Matsubara frequencies is a direct consequence of the fast decay of
propagators $\sim \frac{1}{i\omega_n^2}$ (rather than $\sim \frac{1}{i\omega_n}$, as e.g. in a bare series) and presents a major difference to
diagrammatic algorithms for bare fermionic series formulated in frequency space. At any order at temperature $T/t=0.5$, less than $\sim 4$ frequencies
contribute significantly. 

\begin{figure*}[bth]
\includegraphics[width=0.85\columnwidth]{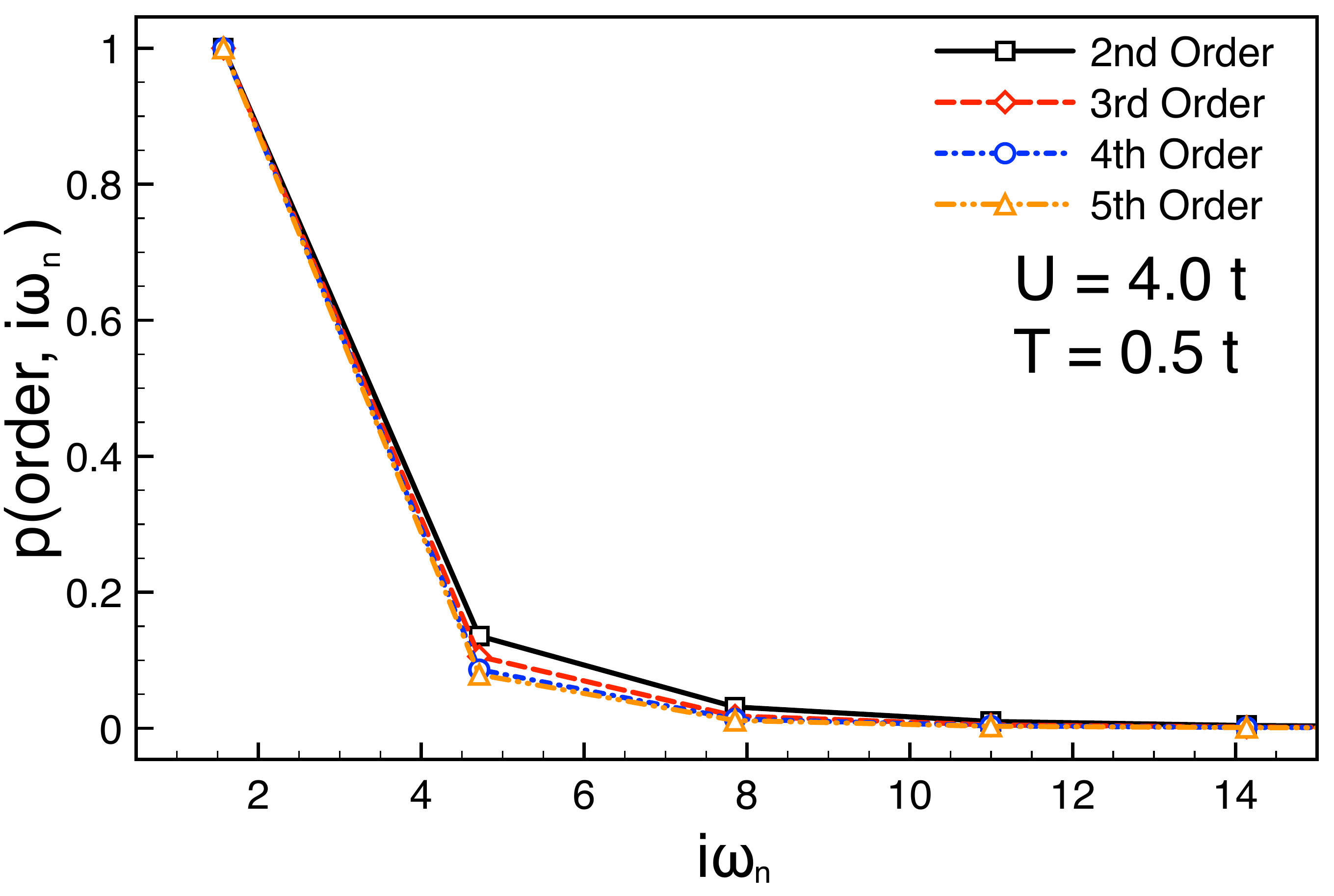}
\includegraphics[width=0.85\columnwidth]{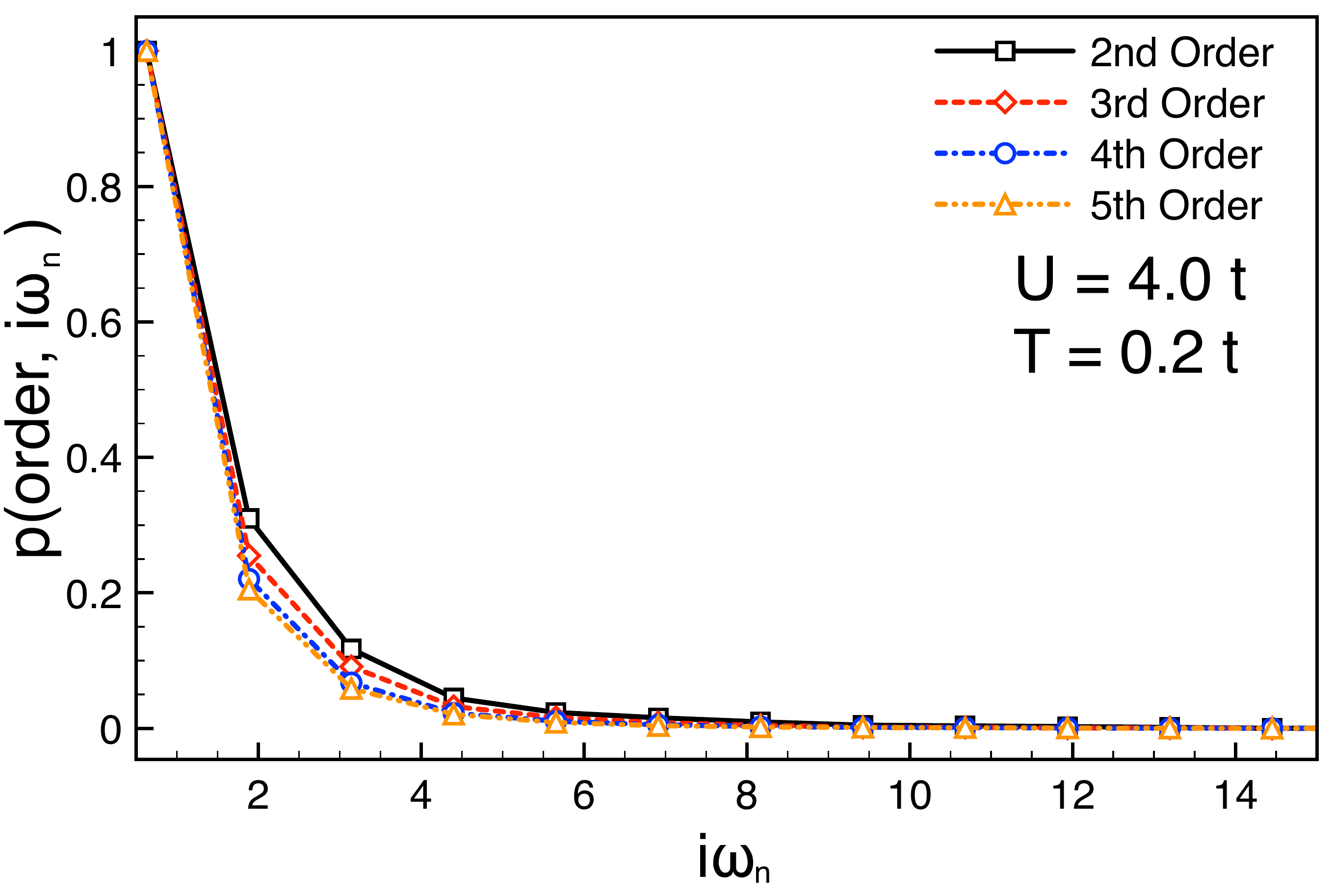}

\includegraphics[width=0.85\columnwidth]{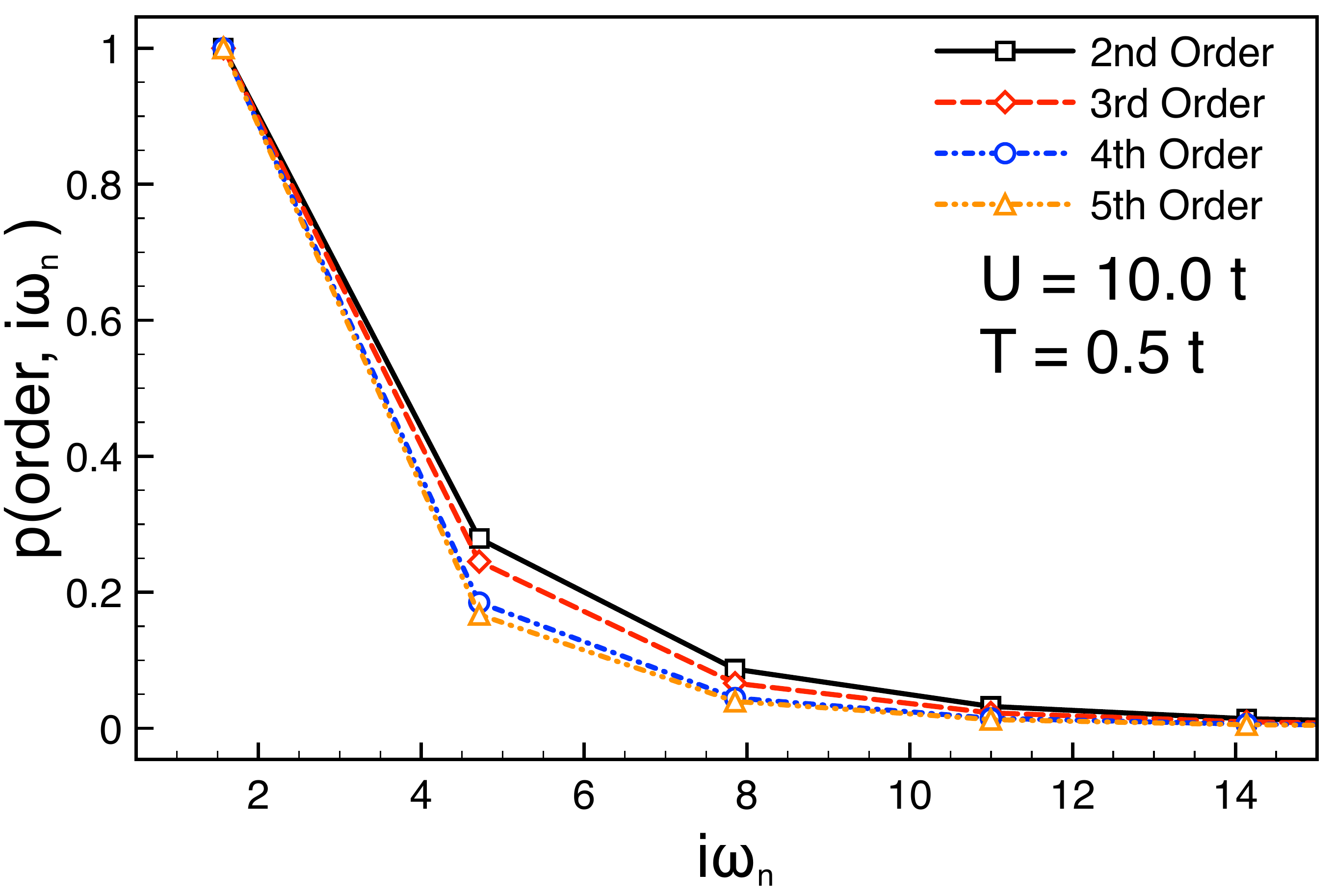}
\includegraphics[width=0.85\columnwidth]{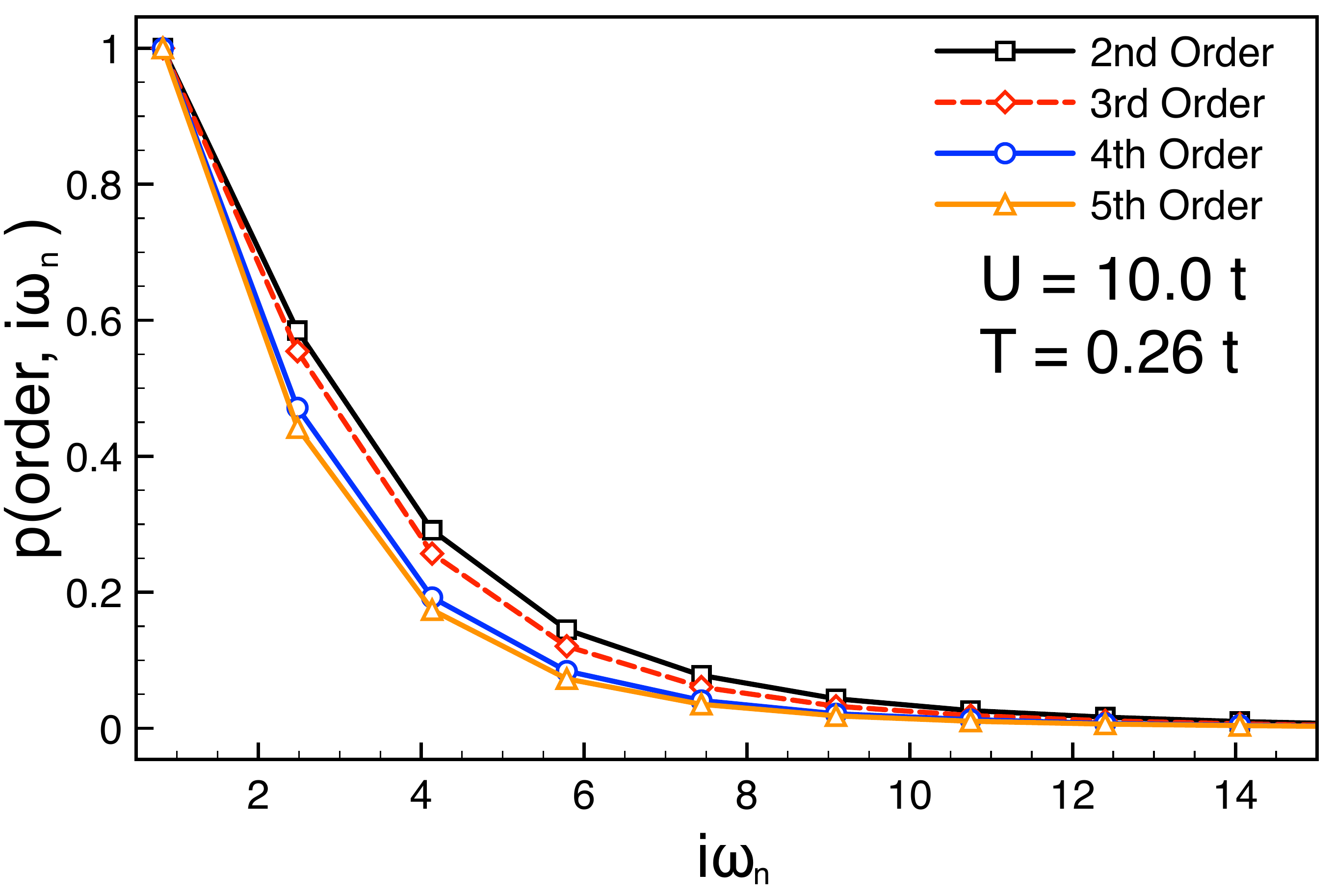}
\caption{Probability that a dual diagram at order $2$, $3$, $4$, or $5$ contains propagators with frequency $i\omega_n$, normalized to the probability of a diagram at the same order containing a propagator at the lowest Matsubara frequency. Data for $U/t = 4.0$ (upper row) and $U/t = 10.0$ (lower row), for
$T/t=0.5$ (left panel) and $T/t=0.2$ and $T/t=0.26$ (right panel).}
\label{fig:OrderMatsubaraT}
\end{figure*}

Decreasing temperature leads to an increase of the number of contributing Matsubara frequencies.
This is shown in Fig.~\ref{fig:OrderMatsubaraT}, where the frequency distribution in the weak and strong coupling regimes at two temperatures $T/t = 0.5$ (left
column) and $T/t = 0.26$ and $T/t = 0.2$ (right column) is plotted. Even at relatively small temperature there is no significant contribution from frequencies higher than
$\omega_n \sim 12$ and the number of required frequencies depends on temperature as $\frac{1}{T}$, implying that the frequency scale of the non-local corrections stays small, so that dual fermion corrections only contribute at low frequencies.

This confinement of the series to low frequencies has practical consequences: it implies that the space needed to be sampled is small,
so that the method converges quickly. It also shows the advantage of sampling diagrams directly in frequency, rather than in imaginary time space. Finally,
because higher order impurity vertices are coupled to more propagator lines, one may expect that their contribution similarly is restricted to low frequencies,
making their computation (and integration into our Monte Carlo method) feasible.

\begin{figure*}[tbh]
\includegraphics[width=0.64\columnwidth]{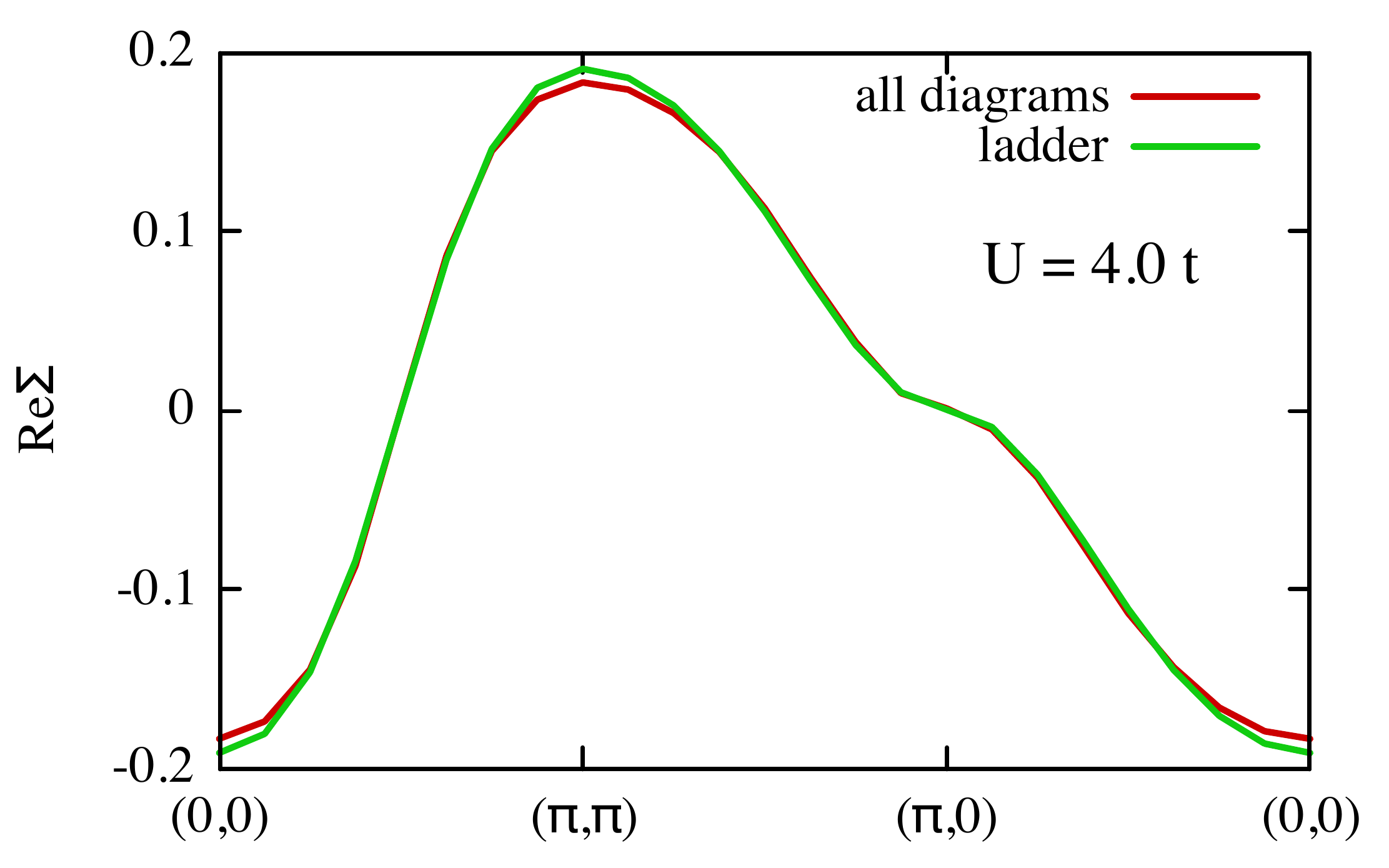}
\includegraphics[width=0.64\columnwidth]{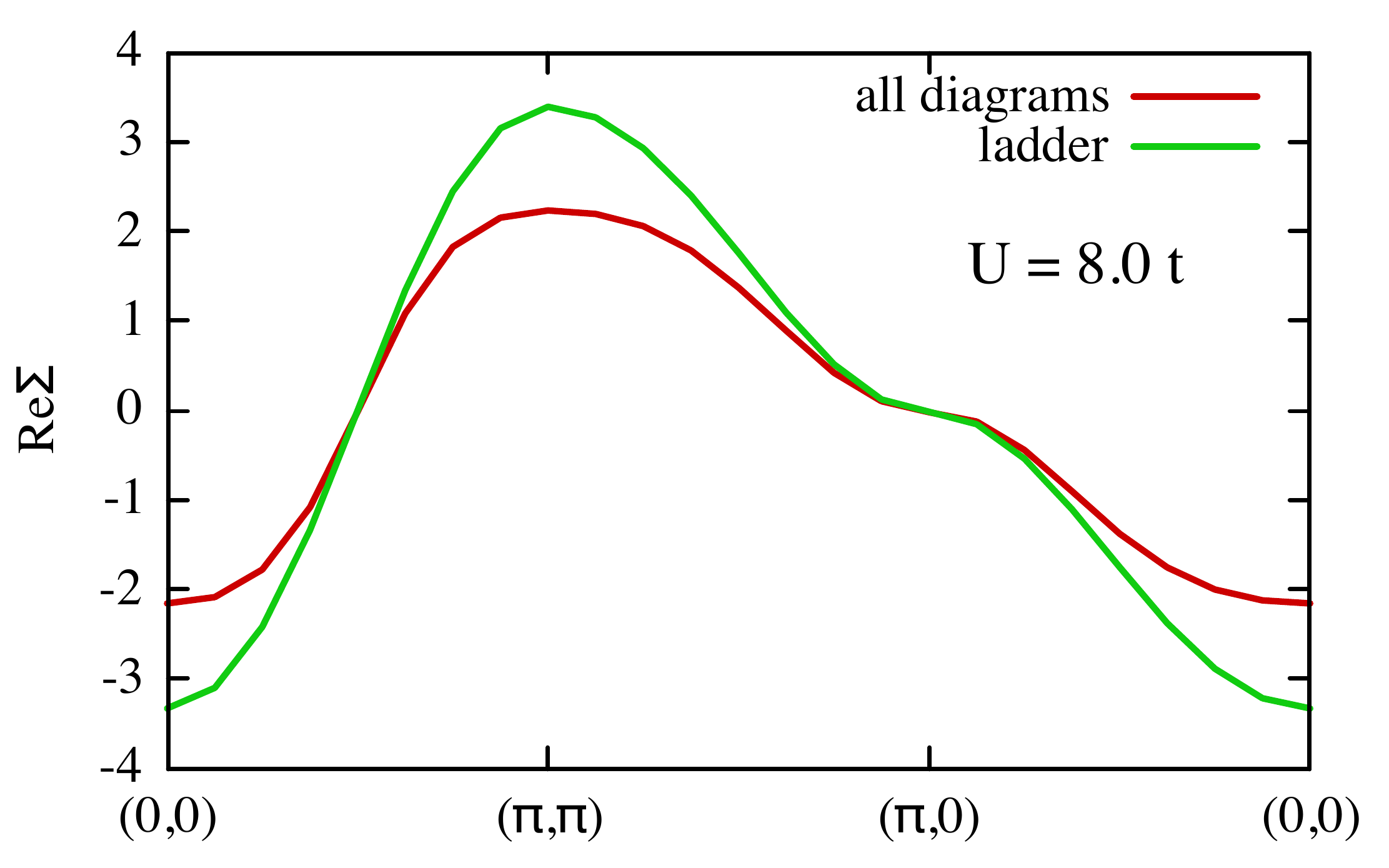}
\includegraphics[width=0.64\columnwidth]{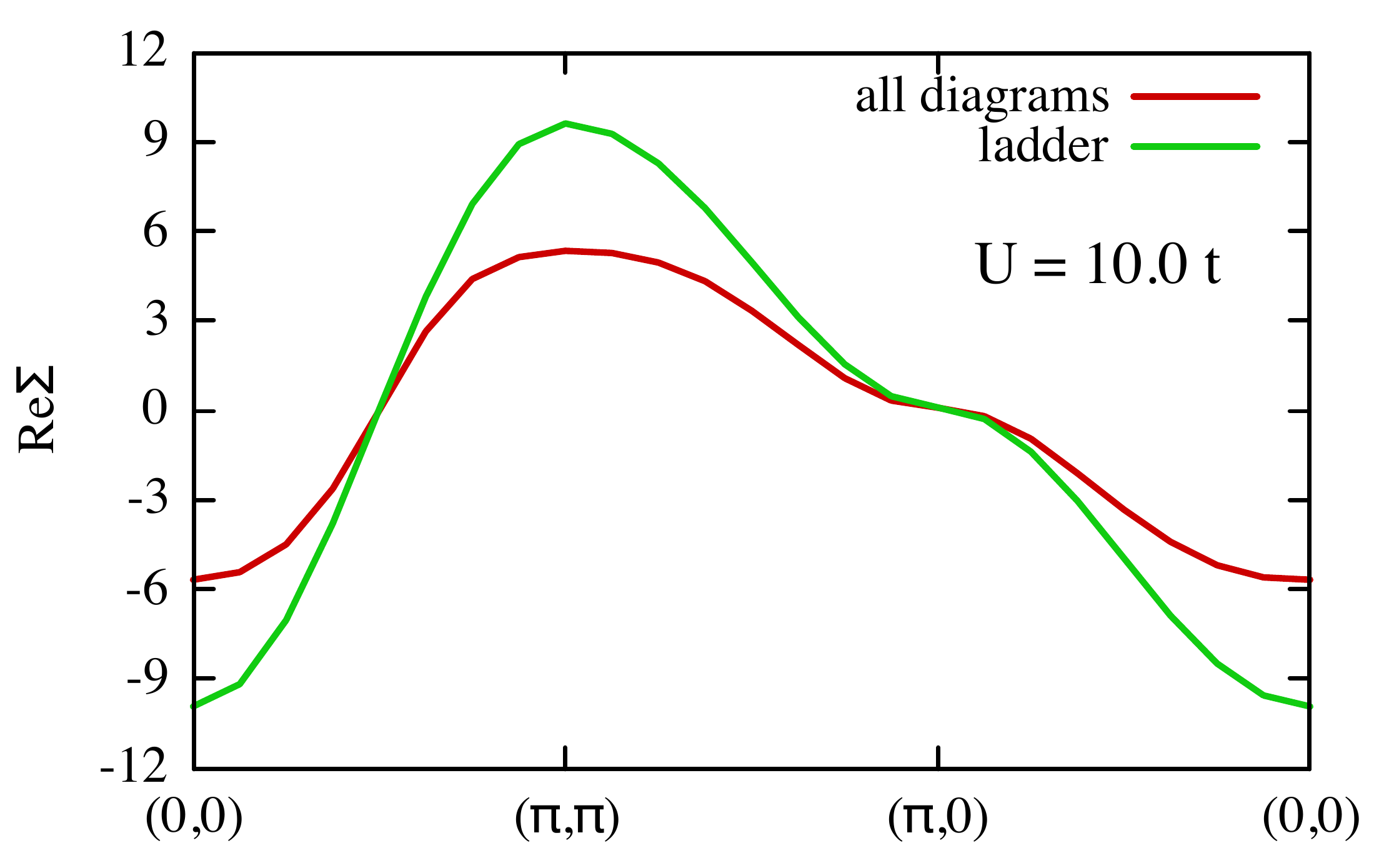}

\includegraphics[width=0.64\columnwidth]{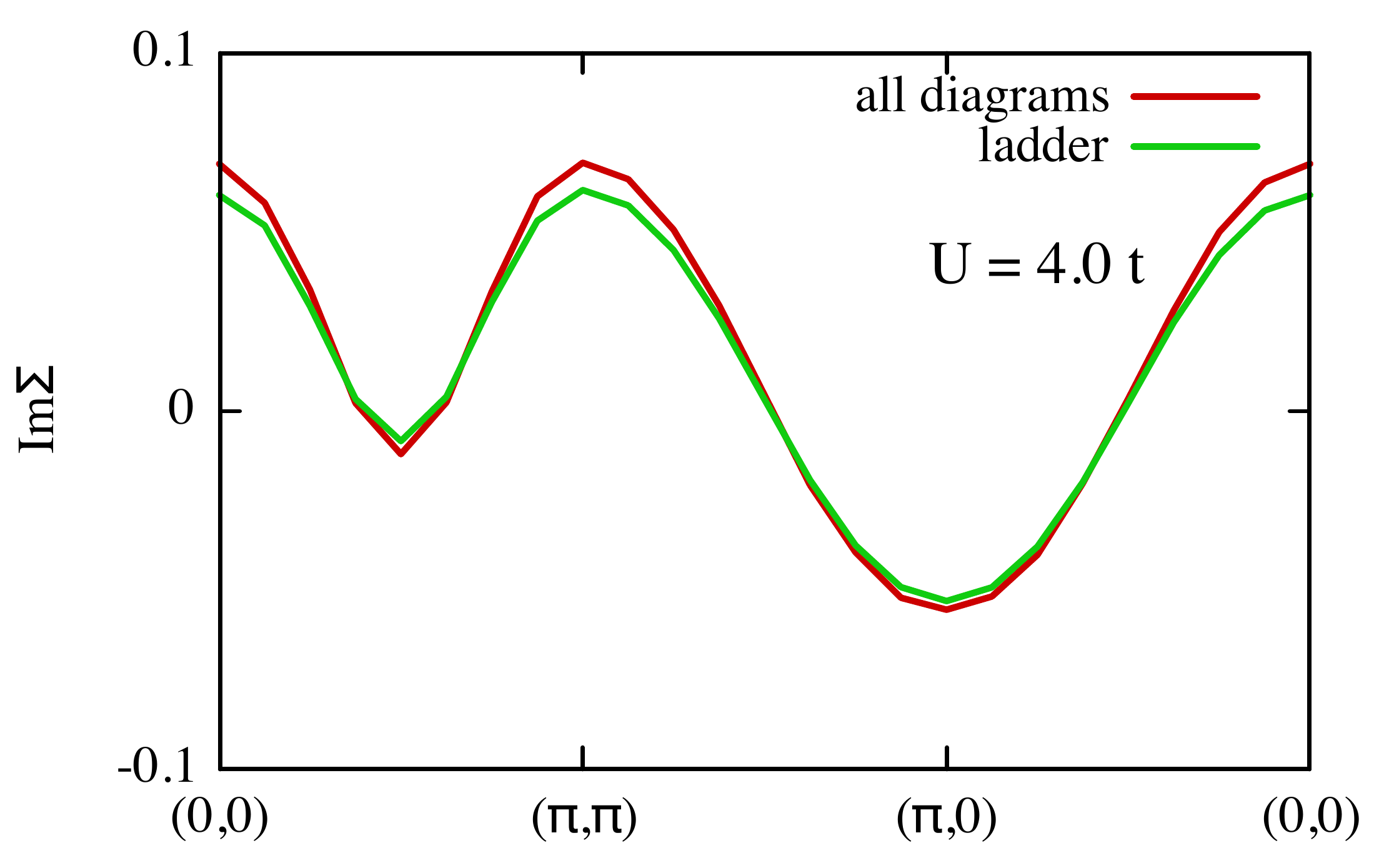}
\includegraphics[width=0.64\columnwidth]{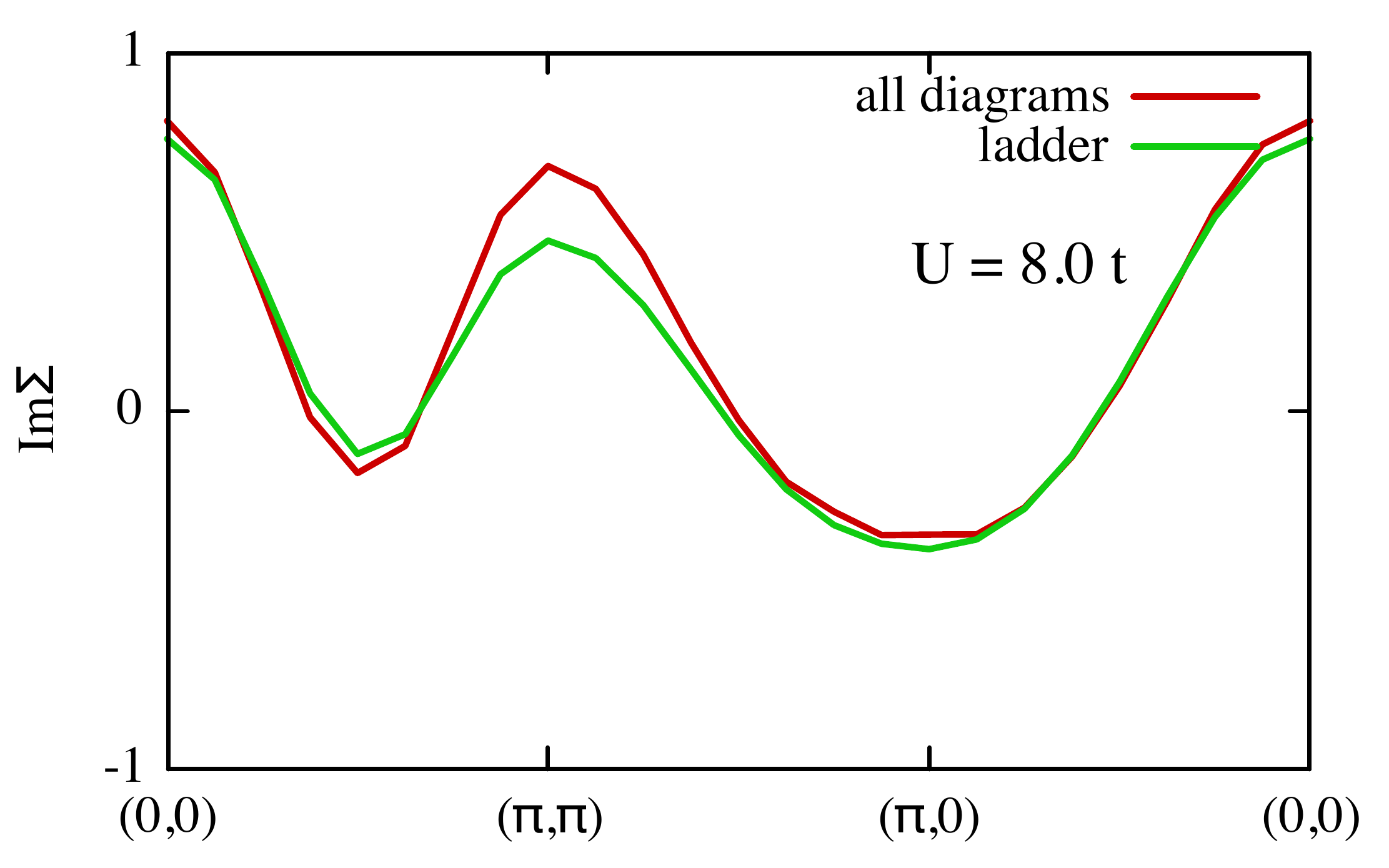}
\includegraphics[width=0.64\columnwidth]{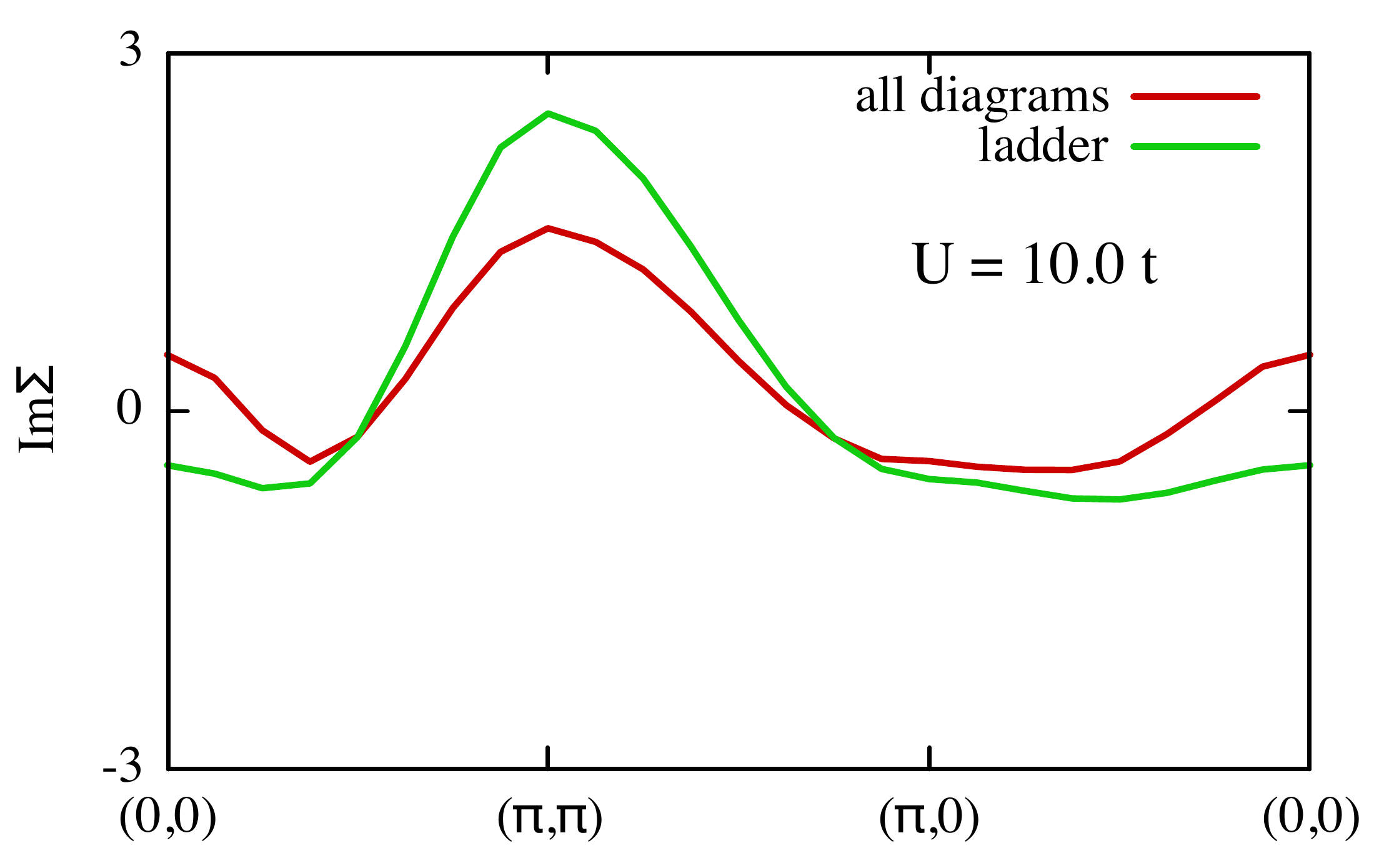}
\caption{Real (upper row) and imaginary (lower row) part of the ladder dual self-energy (green lines) and real part of the self-energy from all diagrams sampled
up to fifth order (red lines) plotted along the main axes of the Brillouin zone at the lowest Matsubara frequency, discretized on a $16\times16$ grid of momenta
and obtained at the coupling strengths $U/t=4$ (left column), $U/t=8$ (middle column) and $U/t=10$ (right column) and at $T/t=0.5$.
}\label{fig:DualSelfenergy}
\end{figure*}
The main output of our simulation are the dual self-energies of Eq.~\ref{eq:sigma_dual}, which are continuous functions of momentum and Matsubara frequency.
These quantities are the input from which lattice Green's functions, energies, and (lattice) self-energies are computed. We plot three examples of this quantity
sampled up to the fifth order in Fig.~\ref{fig:DualSelfenergy} in comparison to the RPA-like ladder summation of the series obtained with the open source
\textit{opendf} code.\cite{opendf} The resulting difference illustrates that in the very weak coupling regime (left panel, $U/t=4, T/t=0.5$), the series is
dominated by a ladder contribution and non-ladder contributions are very small. As the interaction is changed to the intermediate (middle panel, $U/t=8$) and
strong coupling (right panel, $U/t=10$) regime, the non-ladder contributions become substantial, illustrating the importance of our diagrammatic procedure that
samples all possible contributions of two-particle vertex diagrams.

\begin{figure}[tbh]
\includegraphics[width=0.94\columnwidth]{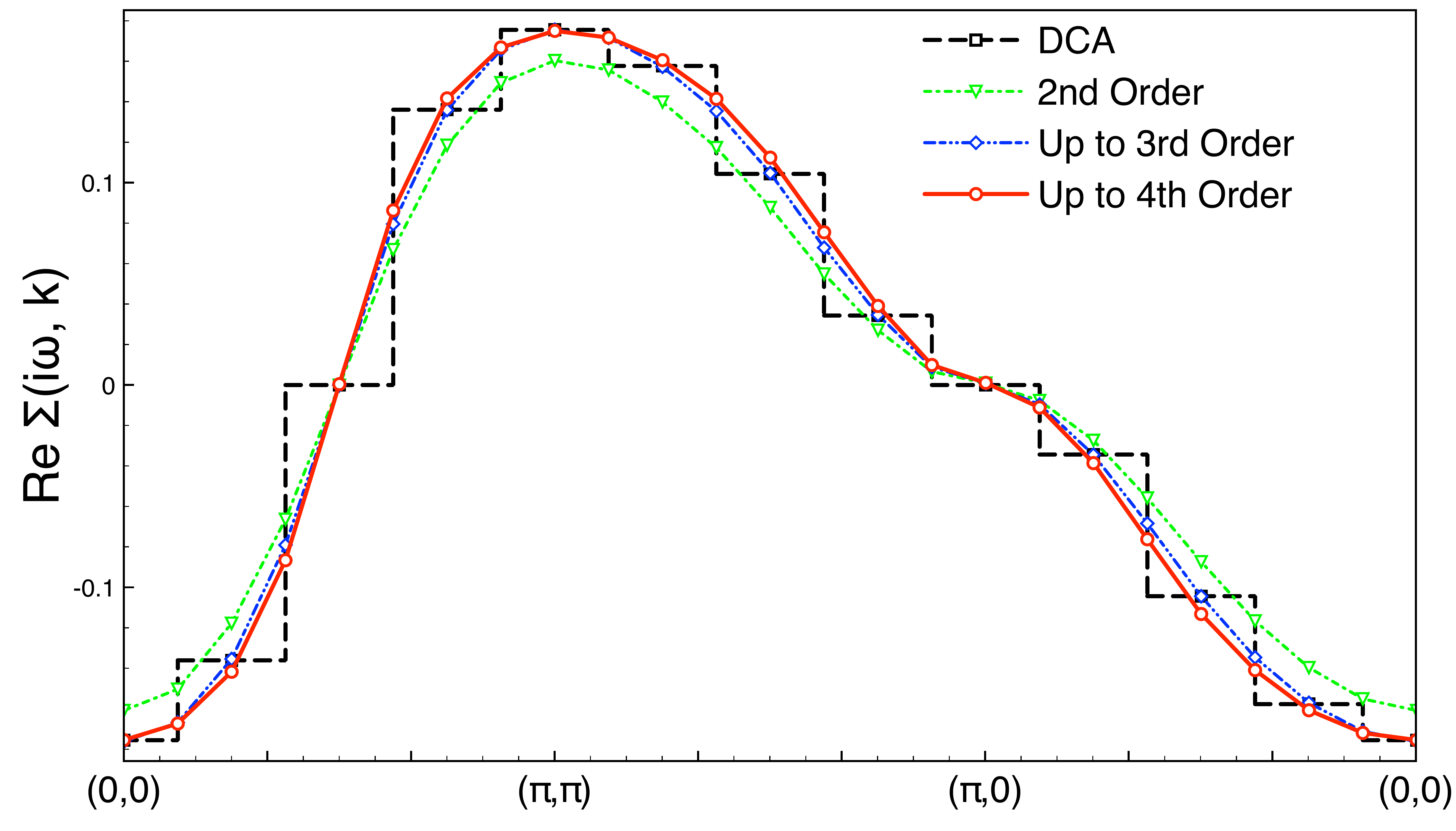}
\caption{Real part of the lattice self-energy plotted along the main axes of the Brillouin zone at the lowest Matsubara frequency, discretized on a $16\times16$ grid of
momenta and obtained for the weak coupling strength $U/t=4$ at $T/t=0.5$, and evaluated up to second order (triangles, green lines), third order (diamonds, blue lines), and 4th order (circles, red lines). Black squares: Comparison to the piecewise constant real self-energy obtained from a dynamical cluster approximation calculation of the same system obtained on a $64$-site cluster.}\label{fig:LatticeSelfenergy}
\end{figure}

The lattice self-energy extracted from the dual self-energy for weak coupling shown in Fig.~\ref{fig:DualSelfenergy} is shown in
Fig.~\ref{fig:LatticeSelfenergy}.
We show results for the real part of the self-energy and results from 2nd, 3rd, and 4th order of diagrammatic dual Fermions. Our results are compared to 64-site dynamical
cluster approximation (DCA) data of Ref.~\onlinecite{LeBlanc15,leblanc:2013} which yield a step-wise constant self-energy. For these parameters, the system is
above any antiferromagnetic ordering temperature (in DMFT or DCA or dual Fermions). The particle hole symmetry is visible as a symmetry between $(0,0)$ and
$(\pi,\pi)$, but antiferromagnetic fluctuations are large and long ranged. Convergence to the DCA solution for these parameters is clearly visible, and the dual
Fermion results provide valuable additional momentum resolution of the self-energy that may in the future allow the resolution of subtle k-space features.

The computational cost of diagrammatic Monte Carlo dual fermion calculations is low compared to other methods for correlated systems. The numerical effort for evaluating the series for parameters examined in this paper ranges from a few minutes for cases where DMFT is accurate to a few hours for cases where the expansion order is high. The computational effort of computing the DMFT vertex functions is strongly dependent on the number of frequencies kept and increases rapidly as $T$ is lowered.

\section{Conclusions}\label{sec:conc}
In conclusion, we have applied a Diagrammatic Monte Carlo method to sample the dual fermion corrections to the dynamical mean field theory and shown results for
the two-dimensional Hubbard model at half filling at small, intermediate and large values of interaction strength. While the method includes all diagram
topologies with two-particle vertices, higher order vertices are neglected. As $T$ is lowered or $U$ changed towards the intermediate interaction regime, non-local contributions beyond second order become relevant. All
corrections are limited to a comparatively narrow frequency range. These contributions contain a substantial part of non-ladder diagrams, illustrating that all
diagram topologies contribute substantially away from weak coupling and critical regimes. The low frequency scale of these diagrams offers the possibility that
three-particle and higher order vertices may be included in future work.
\begin{acknowledgments}
This project was supported by the Simons collaboration on the many-electron problem. We acknowledge helpful discussions with Nikolai Prokof'ev, Boris Svistunov,
Hartmut Hafermann, and Evgeny Kozik. Our diagrammatic Monte Carlo codes are based on the core libraries \cite{ALPSCore} of the open source ALPS \cite{ALPS20}
package, and our DMFT code uses the ALPS implementation \cite{ALPSDMFT} of the continuous-time auxiliary field \cite{Gull08_ctaux,Gull11_submatrix,Gull11_RMP}
(CT-AUX) method with an adaptation of non-equidistant fast Fourier transforms\cite{Staar12,Lin12} to compute vertex functions. 
\end{acknowledgments}

\bibliographystyle{apsrev4-1}
\bibliography{refs.bib}

\end{document}